\documentclass[%
superscriptaddress,
showpacs,
nofootinbib,
amsmath,amssymb,
aps,
prc,
twocolumn,
floatfix ]%
{revtex4-2}

\usepackage{color}
\usepackage{CJK}

\usepackage{graphicx}
\usepackage{dcolumn}
\usepackage{bm}
\usepackage{hyperref}
\usepackage{ifthen}
\renewcommand{\bibinfo}[2]{\ifthenelse{\equal{#1}{title}}{}{#2}}

\allowdisplaybreaks

\begin{document}

\begin{CJK*}{UTF8}{}
\title{Glauber Description of Reaction Cross Sections with Bogoliubov Many-Body Wave Functions}
\CJKfamily{gbsn}
\author{Jie Zhao (赵杰)}%
\affiliation{Engineering Research Center of Advanced Functional Materials and Intelligent Sensing, 
Education Department of Guangxi Zhuang Autonomous Region, 
and Guangxi Key Laboratory of Nuclear Physics and Technology, 
Guangxi Normal University,Guilin 541004, China}
\affiliation{Pengcheng Laboratory, Shenzhen 518055, People's Republic of China}

\date{\today}

\begin{abstract}
I developed a microscopic framework that connects  relativistic nuclear density
functional theory to high-energy reaction observables by incorporating 
many-body Bogoliubov wavefunctions, represented in Pfaffian form,
into the Monte Carlo integration of the Glauber model. 
This approach enables a simultaneous treatment of nuclear pairing correlations 
and multiple-scattering effects beyond the optical limit approximation (OLA).
The method is applied to reaction cross sections
of even-even O, Ne, and Mg isotopes on a $^{12}$C target. 
Complete multiple-scattering calculations systematically 
reduce the cross sections relative to the OLA 
and improves the agreement with measured interaction cross sections. 
Applications to the deformed Ne and Mg isotopes show that 
calculations based on a single intrinsic orientation depend 
strongly on the angle between the nuclear symmetry axis and the beam direction. 
Angular-momentum projection removes the spurious dependence on 
the arbitrarily selected intrinsic orientation and produces smooth 
isotopic systematics for the laboratory-frame cross sections.
This work establishes a microscopic bridge between
relativistic nuclear structure theory and reaction dynamics,
providing a general framework for exploring exotic phenomena 
such as halos and clustering.
\end{abstract}

\maketitle

\end{CJK*}

\bigskip

\section{\label{sec:Introduction}Introduction}
The rapid advancements in accelerator and detection technologies
have enabled the study of quantum objects 
governed by strong interactions through a wide range of 
observables in nucleus-nucleus collisions.
Among the various theoretical frameworks, 
the Glauber model (GM) remains one of the most widely used tools 
for interpreting these data~\cite{Martin1958_PR109-1307,Glauber1970_NPB21-135}.
For example, it serves as a foundation for establishing scaling in 
$pp$, $pA$ and $AA$ collisions and provides realistic initial-state 
conditions for hydrodynamic simulations of the quark-gluon
plasma~\cite{dEnterria2021_ARNPS71-315}.
Furthermore, the GM is extensively employed to analyze 
inclusive cross sections, such as reaction, single-nucleon removal, 
and charge-changing cross sections, revealing exotic nuclear structures 
like halos, skins, and deformations.

The standard Glauber formalism is based on the eikonal approximation, 
assuming that projectile nucleons travel along straight-line trajectories
and undergo multiple, independent sub-collisions 
with nucleons in the target~\cite{dEnterria2021_ARNPS71-315}.
However, the complex multiple-scattering operator is typically 
simplified into a one-body operator, known as the optical limit approximation (OLA).
While convenient, the OLA often washes out crucial 
structural features, leading to an overestimation of the 
total reaction cross sections (RCS)~\cite{Horiuchi2007_PRC75-044607}.

To partially account for the multiple-scattering effects beyond the OLA, 
several approaches have been developed, including the 
nucleon-target formalism~\cite{AbuIbrahim2000_PRC62-034608},
the modified OLA~\cite{Takechi2009_PRC79-061601},
and the use of modified nucleon-nucleon profile functions~\cite{Tran2016_PRC94-064604}.
Additionally, higher-order corrections have been investigated
through expansions in terms of density matrices~\cite{CiofidegliAtti2011_PRC84-025205,
	Hatakeyama2014_JPCS569-012050}.
A particularly effective method for analyzing halo nuclei 
is the few-body Glauber approach, 
in which the core and valence nucleons are treated separately.
In this framework, the OLA is applied to the integration over the 
core-target coordinates, while the integration over the 
valence-nucleon coordinate is performed without any
approximation~\cite{Abu-Ibrahim2003_CPC151-369,Sun2023_CPC288-108726}.
This method has been extensively applied to analyze experimental RCS 
and momentum distributions of fragments, 
allowing the deduction of matter density distributions
in halo candidates~\cite{Kanungo2000_NPA677-171,
	Liu2004_PRC69-034326,Moriguchi2014_NPA929-83,Ogawa1992_NPA543-722,
	Ozawa2008_PRC78-054313,Singh2024_PLB853-138694}.
To fully incorporate the multiple-scattering processes,
in nucleus-nucleus collisions involving projectiles and targets 
with mass numbers $A_{P}$ and $A_{T}$, respectively, 
a $3\times (A_{P}+A_{T}-2)$-dimensional integral must be computed.
In such cases, Monte Carlo integration (MCI) becomes the only viable 
option~\cite{Varga2002_PRC66-034611,Nagahisa2018_PRC97-054614}.

A key ingredient for the MCI is the many-body wavefunction of the colliding nuclei. 
Previous applications have relied on phenomenological
product states consisting of the harmonic-oscillator single-particle 
wave functions~\cite{Nagahisa2018_PRC97-054614,Hatakeyama2019_NPA985-20} 
or the {\it ab initio} wave function obtained from 
the quantum Monte Carlo method~\cite{Varga2002_PRC66-034611,
	Horiuchi2026_PRL136-202501,Horiuchi2026_PRC113-064601}.
At the same time, modern nuclear density functional theories (DFTs)
have achieved remarkable success in describing ground-state
and excited-state properties across the nuclear landscape.
Yet their connection to reaction observables remains largely restricted to 
one-body ground-state densities~\cite{Sharma2016_PRC93-014322,Ahmad2017_PRC95-054601,
	An2024_PLB849-138422,Rashdan2019_PRC99-034604,Zhong2022_SCPMA65-262011,
	Tagami2024_PRC110-064610}.
A consistent treatment that preserves the full many-body structure 
of DFT wave functions in reaction calculations has been missing.

In the present work, I establish such a microscopic connection. 
By projecting relativistic Hartree-Bogoliubov (RHB) wave functions onto 
fixed particle numbers and expressing them in Pfaffian form in coordinate space, 
I embed correlated many-body states directly into the MCI of the GM.
This formulation allows pairing correlations and 
multiple-scattering effects to be treated on an equal footing for the first time.
I first benchmark the method using the nearly spherical $^{16-24}$O isotopes 
and then extend the calculations to the deformed 
$^{18-28}$Ne and $^{20-34}$Mg isotopic chains. 
For the latter systems, I explicitly examine the orientation 
dependence of the RCS and restore rotational symmetry to obtain 
the physical $J=0$ laboratory-frame results. 
The overall agreement with the experimental data 
demonstrates the reliability of the present framework 
for describing high-energy reaction cross sections.
The theoretical framework and method are introduced in Sec.~\ref{sec:model}.
The results are described and discussed in Sec.~\ref{sec:results}. 
Sec.~\ref{sec:summary} contains a summary of the principal results.

\section{\label{sec:model}Theoretical framework}
The high-energy collision of a projectile nucleus with mass
number $A_{P}$ and a target nucleus 
with mass number $A_{T}$ is described  
within the Glauber model~\cite{Glauber1959_LTP1-315}.
Under the adiabatic and eikonal approximations, 
the final state wave function of a projectile and target system 
$\Phi_{f}$ is described by the product of the ground-state wave functions 
of the projectile $\Phi_{0}^{P}$ and the target $\Phi_{0}^{T}$ nuclei, 
and the product of the phase-shift functions of a nucleon-nucleon collision 
$e^{i\chi_{NN}}$ as 
\begin{equation}
	\left| \Phi_{f} \right\rangle = \exp\left\{
		i\sum_{j=1}^{A_{P}}\sum_{k=1}^{A_{T}}
		\chi_{NN}(\bm{b} + \hat{\bm{s}}_{j}^{P} - \hat{\bm{s}}_{k}^{T})
		\right\}
		\left| \Phi_{0}^{P} \Phi_{0}^{T} \right\rangle,
	\label{eq:finallstate}
\end{equation}
where $\bm{b}$ is the impact parameter.
Taking the $z$ direction as the beam direction, 
$\hat{\bm{s}}_{j}^{P}$ ($\hat{\bm{s}}_{k}^{T}$) denotes the two-dimensional 
single-particle coordinate operator projected onto the $x$-$y$ 
plane of the $j$th ($k$th) nucleon from the center of mass of the projectile (target).
With this approximation, we only need to evaluate the optical phase-shift function 
\begin{equation}
	e^{i\chi(\bm{b})} = 
	\left\langle \Phi_{0}^{P} \Phi_{0}^{T} 
	\left| \prod_{j=1}^{A_{P}} \prod_{k=1}^{A_{T}} 
	\left[ 1- \Gamma_{NN}(\bm{b} + \hat{\bm{s}}_{j}^{P} - \hat{\bm{s}}_{k}^{T}) \right]
	\right| \Phi_{0}^{P} \Phi_{0}^{T} \right\rangle,
	\label{eq:phase}
\end{equation}
where the profile function is defined as
\begin{equation}
	\Gamma_{NN}(\bm{b}) = 1- e^{i\chi_{NN}(\bm{b})}.
	\label{eq:profile}
\end{equation}
The total RCS is evaluated by
\begin{equation}
	\sigma_{R} = \int d\bm{b} P(\bm{b}),
	\label{eq:sigma}
\end{equation}
where $P(\bm{b})$ is the reaction probability
\begin{equation}
	P(\bm{b}) = 1 - \left| e^{i\chi(\bm{b})} \right|^{2}.
	\label{eq:probability}
\end{equation}
The profile function is parametrized as 
\begin{equation}
	\Gamma_{NN}(\bm{b}) = \frac{1-i\alpha_{NN}}{4\pi\beta_{NN}} 
	\sigma_{NN}^{\rm{tot}} \exp\left[-\frac{\bm{b}^{2}}{2\beta_{NN}} \right],
	\label{eq:profile2}
\end{equation}
where $\sigma_{NN}^{\rm{tot}}$, $\alpha_{NN}$, and $\beta_{NN}$ are the 
total nucleon-nucleon cross section, the ratio between the real and imaginary
parts of the scattering amplitude at the forward angle, 
and the slope parameter, respectively.
The parameter sets for various incident energies listed in 
Ref.~\cite{AbuIbrahim2008_PRC77-034607} for proton-proton (pp)
and proton-neutron (pn) are employed. 
The nn (np) are taken to be the same as pp (pn).
Introducing the $A$-body density 
\begin{equation}
	\rho_{A}(\bm{R}) = 
	\left\langle \Phi_{0} \left| \prod_{i=1}^{A} \delta(\hat{\bm{r}}_{i}-\bm{r}_{i})
	\right| \Phi_{0} \right\rangle,
	\label{eq:density}
\end{equation}
where $\bm{R} \equiv (\bm{r}_{1},\cdots,\bm{r}_{A}$) and $\hat{\bm{r}}_{i}$
is the single-particle coordinate operator of the $i$th nucleon from the origin.
The complete Glauber amplitude can be written as 
\begin{equation}
\begin{aligned}
	e^{i\chi(\bm{b})} =& \int \cdots \int 
	\left( \prod_{j=1}^{A_{P}} d\bm{r}_{j}^{P} \right) 
	\left( \prod_{k=1}^{A_{T}} d\bm{r}_{k}^{T} \right) \\
  & \times \rho_{A_{P}}^{P}(\bm{R}^{P})
    \rho_{A_{T}}^{T}(\bm{R}^{T}) \\
  & \times \prod_{j=1}^{A_{P}} \prod_{k=1}^{A_{T}} 
	\left[ 1- \Gamma_{NN}(\bm{b} + \tilde{\bm{s}}_{j}^{P} - \tilde{\bm{s}}_{k}^{T}) \right],
\end{aligned}
\label{eq:phase_mb}
\end{equation}
where $\tilde{\bm{s}}_{j}^{P} $($\tilde{\bm{s}}_{k}^{T}$) denotes the $xy$ component 
of the $j$th ($k$th) single-particle coordinate from the center-of-mass coordinate 
of the projectile (target) ($\tilde{\bm{r}}_{j} = \bm{r}_{j} - \bm{X}$ with 
$\bm{X}=\frac{1}{A} \sum_{j=1}^{A} \bm{r}_{j}$).
Eq.~\ref{eq:phase_mb} involves $3(A_{P}+A_{T})$-dimensional integration.
Following Refs.~\cite{Varga2002_PRC66-034611,Nagahisa2018_PRC97-054614,
	Kaki2012_PRC86-044601,Horiuchi2026_PRL136-202501,Horiuchi2026_PRC113-064601}
I employ MCI to evaluate this multidimensional integration 
and $\rho_{A_{P}}^{P}\rho_{A_{T}}^{T}$ 
is used as the guiding function of the Metropolis iterative algorithm.

For the ground-state wave functions of the 
projectile $\Phi_{0}^{P}$ and target $\Phi_{0}^{T}$,
products of single-particle wave functions 
generated from phenomenological potential~\cite{Nagahisa2018_PRC97-054614}
or wave functions obtained by the QMC method~\cite{Varga2002_PRC66-034611,
	Kaki2012_PRC86-044601,Horiuchi2026_PRL136-202501,Horiuchi2026_PRC113-064601}
are used in previous works.
In the present work, I use the many-body Bogoliubov wave function 
generated from the RHB model where the RHB equation
\begin{equation}
	\label{eq:rhb}
	\left(\begin{array}{cc} h_{D}-\lambda & \Delta \\ -\Delta^{*} & -h_{D}+\lambda \end{array}\right)
	\left(\begin{array}{c} U_{k} \\ V_{k} \end{array}\right)
	= E_{k}\left(\begin{array}{c} U_{k} \\ V_{k} \end{array}\right),
\end{equation}
is solved by expanding the Dirac spinors in a harmonic oscillator (HO) basis.
$E_{k}$ is the quasiparticle energy, 
$\left( U_k, \ V_{k} \right)^\mathrm{T}$ is the quasiparticle wave function,
$\lambda$ is the chemical potential 
and $h_{D}$ is the single-particle Dirac Hamiltonian,
$\Delta$ is the pairing potential.
The corresponding Bogoliubov many-body wave functions, 
in the Thouless representation~\cite{Ring1980_NMBP}, are given by 
\begin{equation}
\label{eq:thouless}
\left| \Phi_{0} \right\rangle = \mathcal{N} \exp{\left( 
	\frac{1}{2} \sum_{kk^{\prime}} Z_{kk^{\prime}} 
	c_{k}^{\dagger} c_{k^{\prime}}^{\dagger} \left| 0 \right\rangle
	\right)},
\end{equation}
where the skew-symmetric matrices $Z = (VU^{-1})^{*}$
and $\left| 0 \right\rangle$ is the true vacuum.
$c_{k}^{\dagger}$ and $c_{k}$ represent the single 
particle creation and annihilation operators.
The coordinate-space amplitude for an even number of $A=2n$ nucleons
can be expressed as
\begin{equation}
\label{eq:coor}
	\Phi_{0}(\bm{R}) 
	= \left\langle 0 \left| 
 	\hat{\psi}(\bm{r}_{A}) \cdots \hat{\psi}(\bm{r}_{1})
	\right| \Phi_{0} \right\rangle,
\end{equation}
where 
\begin{equation}
\label{eq:fieldoperator}
	\hat{\psi}(\bm{r}) = \sum_{i} \phi_{i}^{*} (\bm{r}) c_{i},
\end{equation}
is the field operator.
The Bogoliubov vacuum does not have fixed particle number.
By projecting the wave functions Eq.~\ref{eq:thouless} 
onto the $A$ particle sector, I obtain
\begin{equation}
\begin{aligned}
	\Phi_{0}(\bm{R})
	=& \frac{\mathcal{N}}{2^{n}n!}
	\sum_{\{i\}} \phi_{i_{1}}^{*}(\bm{r}_{1}) \cdots \phi_{i_{2n}}^{*}(\bm{r}_{2n}) \\
	& \times 
	\sum_{k} Z_{k_{1}k_{2}} \cdots Z_{k_{2n-1}k_{2n}} \\
	& \times 
	\left\langle 0 \left| 
	c_{i_{2n}} \cdots c_{i_{1}}
	c^{\dagger}_{k_{1}} \cdots c^{\dagger}_{k_{2n}}
	\right| 0 \right\rangle.
\end{aligned}
\end{equation}
Applying the fermionic Wick theorem yields
\begin{equation}
\label{eq:coor}
	\Phi_{0}(\bm{R}) = \mathcal{N} \rm{Pf}
	\left[ \mathcal{F}(\bm{r}_{i}, \bm{r}_{j}) \right]_{i,j=1}^{2n},
\end{equation}
where
\begin{equation}
\label{eq:cele}
	\mathcal{F}(\bm{r}_{i}, \bm{r}_{j}) = 
	\sum_{kk^{\prime}} \phi_{k}^{*}(\bm{r}_{i}) Z_{kk^{\prime}}
	\phi_{k^{\prime}}^{*}(\bm{r}_{j}),
\end{equation}
and $\phi_{k}(\bm{r})$ is the HO single-particle wave function.
This representation encodes two-body correlations beyond a simple Slater determinant.
For a deformed nucleus, the intrinsic Bogoliubov vacuum does not have definite 
angular momentum. The laboratory-frame state with angular momentum $J$ and 
projection $M$ is~\cite{Ring1980}
\begin{equation}
	\label{eq:proj}
	\left| \Phi^{JM}_{0} (Z) \right\rangle 
	= \sum_{K} f^{JK} \hat{P}^{J}_{MK} \left| \Phi_{0} (Z) \right\rangle,
\end{equation}
where
\begin{equation}
	\label{eq:amp3d}
	\hat{P}^{J}_{MK} = \frac{2J+1}{8\pi^{2}} \int \mathcal{D}_{MK}^{J*}(\Omega)
	\hat{R}(\Omega) d\Omega,
\end{equation} 
and $\hat{R}(\Omega)$ is the rotation operator.
$\Omega\equiv(\alpha,\beta,\gamma)$ is the Euler angle.
Note $\hat{R}(\Omega) \left| \Phi_{0} (Z) \right\rangle 
= \left| \Phi_{0}[Z^{(\Omega)}] \right\rangle$ with 
$Z^{(\Omega)} = DZD^{T}$ and $D_{ji}(\Omega) 
= \langle j | \hat{R}(\Omega) | i \rangle$,
I can write 
\begin{equation}
	\label{eq:proj2}
	\left| \Phi^{JM}_{0} (Z) \right\rangle 
	= \int d\Omega w^{JM}(\Omega) \left| \Phi_{0} [Z^{(\Omega)}] \right\rangle,
\end{equation}
with
\begin{equation}
	\label{eq:defw}
	w^{JM}(\Omega) = \frac{2J+1}{8\pi^{2}}
	\sum_{K} f^{JK} \mathcal{D}_{MK}^{J*}(\Omega).
\end{equation}
Because $[\hat{N},\hat{R}(\Omega)]=0$,
following the same derivation of Eq.~\ref{eq:coor}, 
I obtain 
\begin{equation}
	\label{eq:amp_coor}
	\Phi_{0}^{JM}(\bm{R}) = \mathcal{N}^{JM} \int d\Omega w^{JM} (\Omega)\rm{Pf}
	\left[ \mathcal{F}^{(\Omega)}(\bm{r}_{i}, \bm{r}_{j}) \right]_{i,j=1}^{2n},
\end{equation}
with
\begin{equation}
	\label{eq:amp_cele}
	\mathcal{F}^{(\Omega)}(\bm{r}_{i}, \bm{r}_{j}) = 
	\sum_{kk^{\prime}} \phi_{k}^{*}(\bm{r}_{i}) Z_{kk^{\prime}}^{(\Omega)}
	\phi_{k^{\prime}}^{*}(\bm{r}_{j}).
\end{equation}
For the axially symmetric and reflection-symmetric even-even nuclei considered
in the present work, the integration over the Euler angle $\gamma$
can be carried out analytically. 
The exact projected coordinate-space wave function consequently involves 
a two-dimensional (2D) integration over $\alpha$ and $\beta$.
Direct evaluation of this 2D calculation
is computationally demanding. 
I therefore employ a one-dimensional (1D) axial projection approximation in 
which only rotations about the $y$ axis are retained.
The approximate projected coordinate-space amplitude is written as 
\begin{equation}
	\label{eq:amp_1d}
	\tilde{\Phi}_{0}^{J}(\bm{R}) = \tilde{\mathcal{N}}^{J} 
	\int_{0}^{\pi} d\beta \tilde{w}^{J} (\beta)\rm{Pf}
	\left[ \mathcal{F}^{(\beta)}(\bm{r}_{i}, \bm{r}_{j}) \right]_{i,j=1}^{2n},
\end{equation}
with
\begin{equation}
	\label{eq:weight_1d}
	\tilde{w}^{J} (\beta) = \frac{2J+1}{2} \sin\beta d_{00}^{J}(\beta).
\end{equation}
To avoid singularity when calculate $Z$ for the case of vanishing pairing, 
a regularization factor $\eta = 10^{-3}$ is added to $\Delta_{i\bar{i}}$. 
Since the Glauber theory is formulated within a non-relativistic kinematic 
framework and the resulting profile function Eq.~\ref{eq:profile2} 
is therefore non-relativistic, as a first attempt to apply a relativistic 
microscopic approach to reaction observables, I neglect the contributions 
from the small components and retain only the large components to construct 
the coordinate-space many-body wave function Eq.~\ref{eq:coor}.
One should emphasize that, although several previous works have 
attempted to connect relativistic DFT to reaction observables via 
the Glauber model, these studies are limited to the OLA and require 
only the one-body density from relativistic 
DFT~\cite{An2024_PLB849-138422,Ahmad2017_PRC95-054601,Das2023_NPA1037-122703}.

The relativistic energy density functionals (REDFs) are rooted in 
quantum hadrodynamics (QHD), in which a nucleus is described 
as a system of Dirac nucleons interacting via meson exchange 
through an effective Lagrangian 
(see, e.g., Ref.~\cite{Niksic2011_PPNP66-519}). 
In the mean-field approximation, the theory can be reformulated in terms of 
equivalent local interactions with density-dependent couplings. 
The parameters of a given REDF are constrained by nuclear matter properties 
and further fine-tuned to experimental data of finite nuclei. 
In the present work, I employ the DD-PC1 parameter set~\cite{Niksic2008_PRC77-034302}.
For the pairing channel, I adopt a separable finite-range interaction 
adjusted to reproduce the density dependence of the pairing gap in 
symmetric nuclear matter obtained with the Gogny
force~\cite{Tian2009_PLB676-44,Tian2009_PRC80-024313,Niksic2010_PRC81-054318}. 
A fully consistent derivation of both the nuclear structure input and the reaction profile function within a unified relativistic microscopic 
framework remains highly nontrivial. 
Therefore, the nucleon-nucleon profile function in Eq. (6) is taken from 
empirical nucleon-nucleon scattering data~\cite{AbuIbrahim2008_PRC77-034607}.
Calculations are performed in a basis with 10 oscillator shells.
As in Ref.~\cite{Nagahisa2018_PRC97-054614}, Coulomb breakup contributions 
are neglected, as they are expected to be small for light systems.
I emphasize that the present approach is not a fully microscopic 
or ab initio framework in which the nuclear structure and reaction 
ingredients are derived from a common underlying interaction. 
The term ``microscopic" is used here to indicate that the self-consistent 
many-body Bogoliubov wave functions are incorporated directly into the 
complete Glauber multiple-scattering calculation, 
rather than entering the reaction model only through their one-body densities.

\section{\label{sec:results}Results and discussions}

Spatial configurations in the $A$-body Cartesian coordinate space  
$(\bm{r}_{1},\cdots,\bm{r}_{A})$ are generated via a Metropolis random walk 
with a step size of $\delta = 1.0$ fm.
The resulting $10^{6}$ MC configurations, sampled according to the probability distribution 
of the guiding function are used to perform the multidimensional integration 
over projectile and target coordinates.
The 2D integration over $\bm{b}$ is performed 
using a 21st-order Gauss-Kronrod quadrature.
To evaluate Eq.~\ref{eq:amp_1d} for axially deformed nuclei, 
the integration over $\beta$ is performed using
a $14$-point Gauss-Legendre quadrature.

\begin{figure}[htb]
	\begin{center}
		\includegraphics[width=0.45\textwidth]{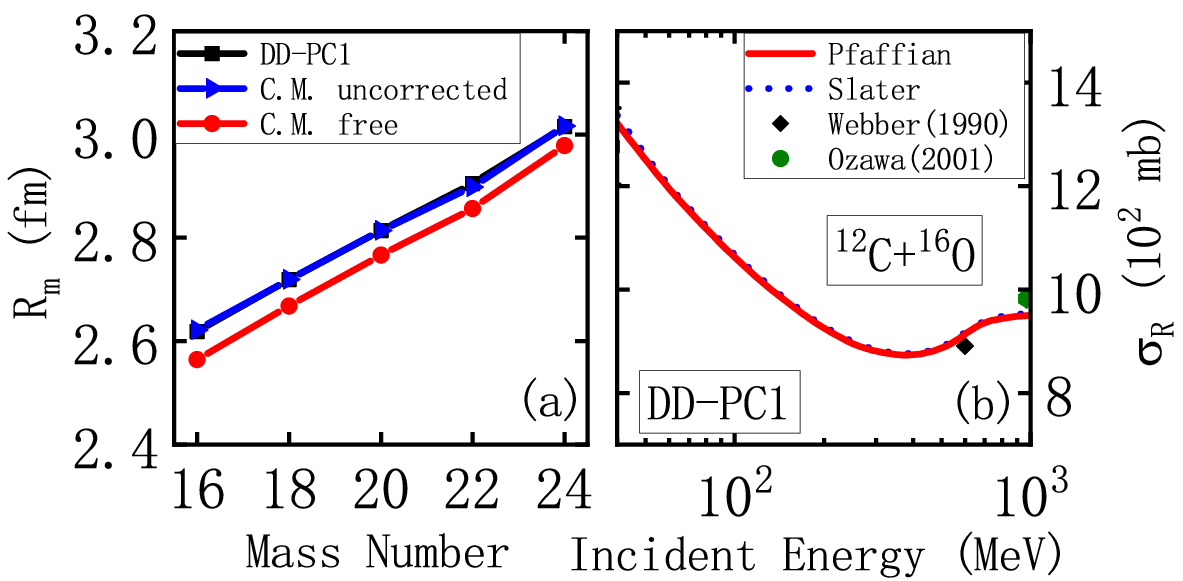}
	\end{center}
	\caption{(Color online) 
	(a) The center-of-mass (C.M.) uncorrected (blue triangles) and the C.M. corrected (red dots)  
	root-mean-square (RMS) radii of $^{16-24}$O calculated from the MCI
	with configurations generated from the Pfaffian wave functions.
	The RMS radii from the RHB calculations with DD-PC1 (black squares) 
	are also shown for comparison. 
	(b) Energy dependence of the total reaction cross sections of the $^{12}$C+$^{16}$O collision.
	The data are taken from Ref.~\cite{Webber1990_PRC41-520} (black diamond) and Ref.~\cite{Ozawa2001_NPA693-32} (Olive dot).
	}
	\label{fig:rmsR}
\end{figure}

\begin{figure}[htb]
	\begin{center}
		\includegraphics[width=0.45\textwidth]{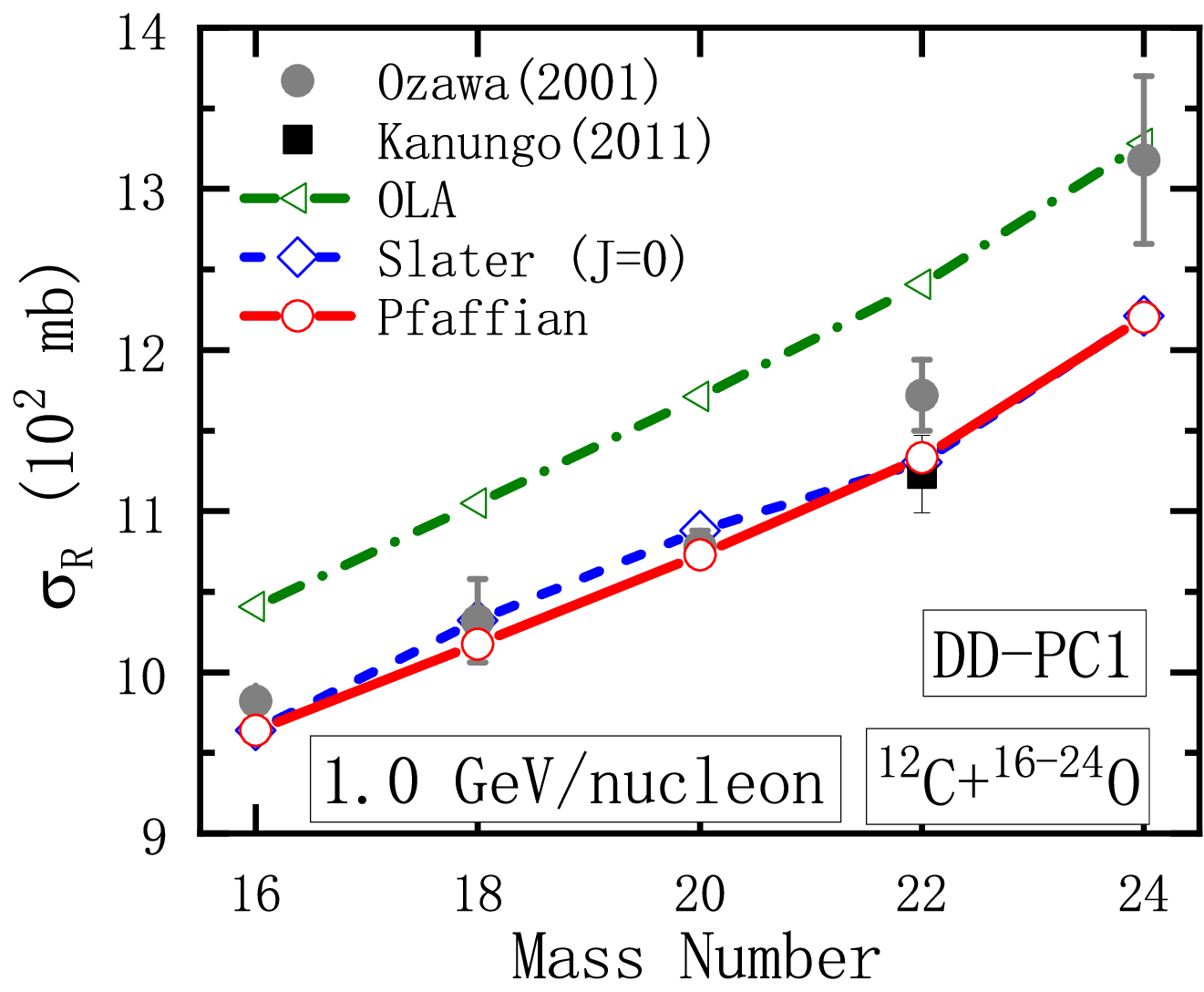}
	\end{center}
	\caption{(Color online) 
	Reaction cross sections of $^{16-24}$O isotopes on a $^{12}$C target at 1.0 GeV/nucleon.
	The OLA results calculated using the RHB one-body densities are shown by olive left-pointing triangles.
	The MCI results obtained using configurations sampled from the Pfaffian and the
	(J=0) angular-momentum-projected  Slater-determinant wave functions are shown by red circles and blue diamonds, respectively.
	The data (interaction cross sections) are taken from Ref.~\cite{Ozawa2001_NPA693-32} (gray dots) and Ref.~\cite{Kanungo2011_PRC6-061304} (black squares).
	}
	\label{fig:Ocs}
\end{figure}

\begin{figure}[htb]
	\begin{center}
		\includegraphics[width=0.45\textwidth]{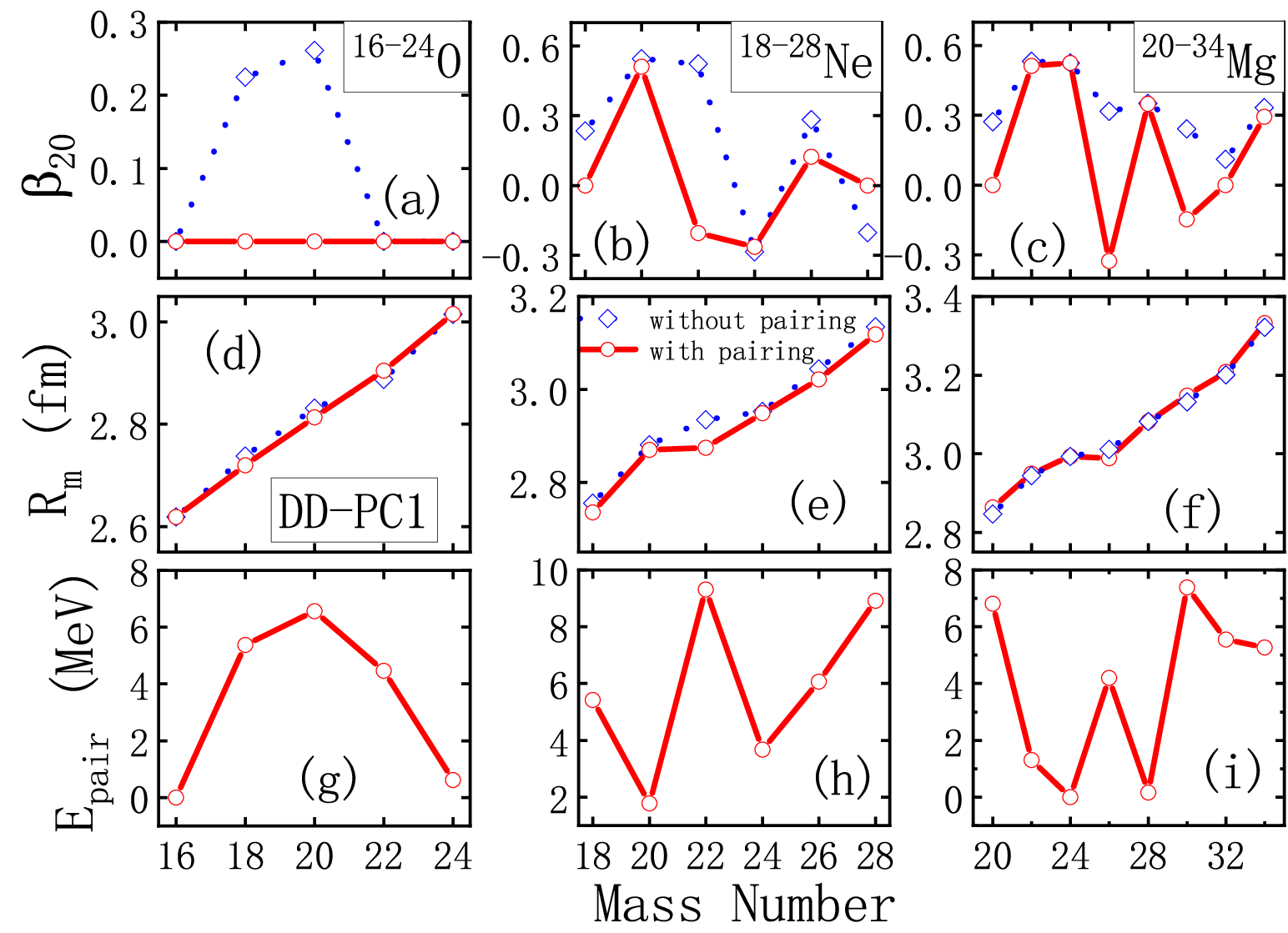}
	\end{center}
	\caption{(Color online) 
	Ground-state properties of the even-even $^{16-24}$O, $^{18-28}$Ne, and $^{20-34}$Mg isotopes calculated with DD-PC1.
	Panels (a)-(c) show the quadrupole deformation $\beta_{20}$, panels (d)-(f) the matter RMS radius $R_{m}$, and panels (g)-(i) the pairing energy. 
	Results obtained with and without pairing correlations are distinguished by red circles with solid lines and blue diamonds with dotted lines, respectively.
	}
	\label{fig:gsIso}
\end{figure}

\begin{figure}[htb]
	\begin{center}
		\includegraphics[width=0.45\textwidth]{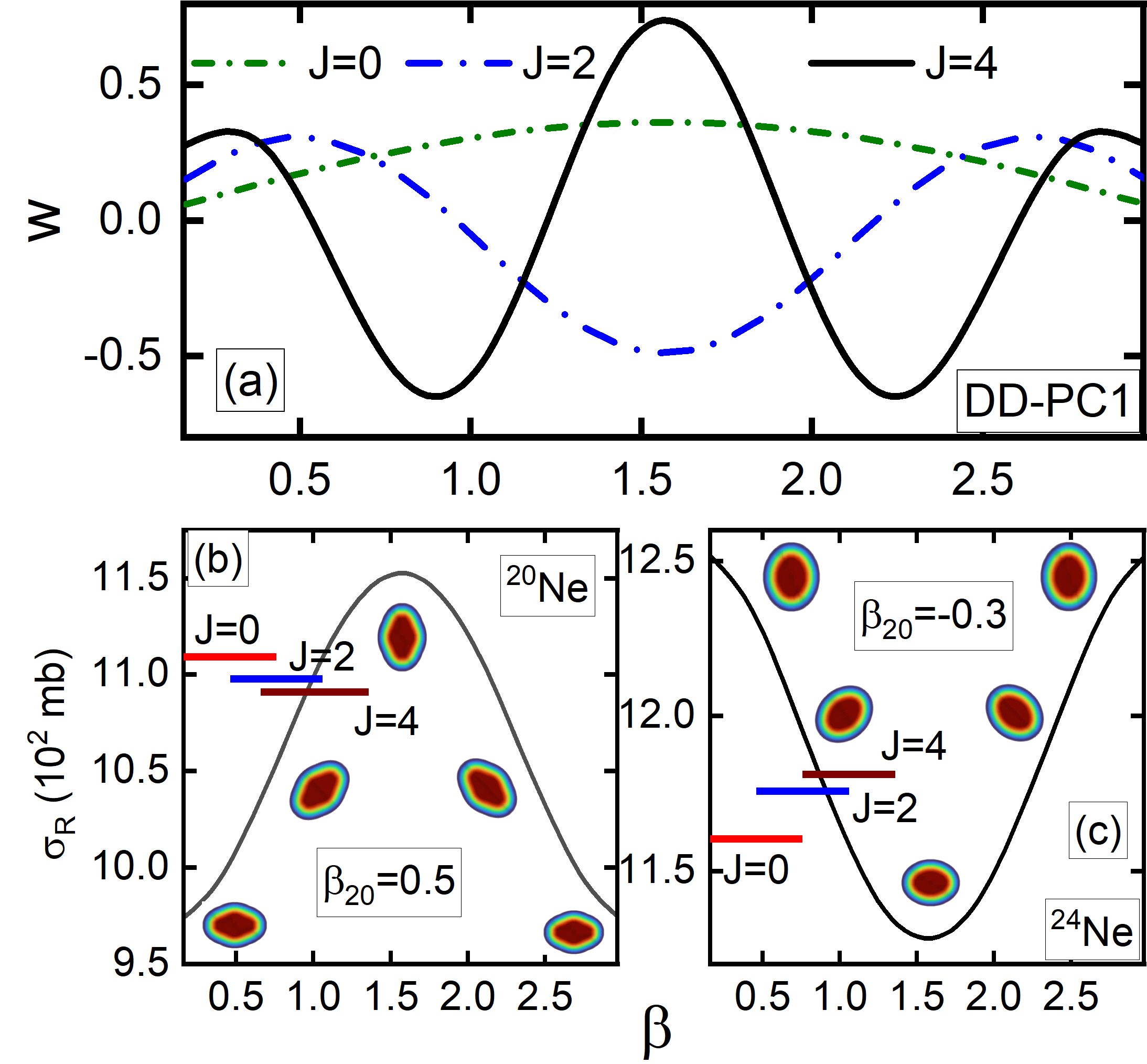}
	\end{center}
	\caption{(Color online) 
		Orientation and angular-momentum-projection effects on reaction cross sections. Panel (a) shows the angular weights entering the $J=0$, 2, and 4 projections.
		Panels (b) and (c) show the reaction cross section as a function of the angle 
		between the intrinsic symmetry axis 
		and the beam for the prolate $\beta_{20}\approx0.5$ configuration of $^{20}$Ne and the oblate
		$\beta_{20}\approx-0.3$ configuration of $^{24}$Ne, respectively.
		The horizontal colored lines indicate 
		the corresponding angular-momentum-projected values. 
		Representative intrinsic density distributions
		are included to illustrate the collision geometry.
	}
	\label{fig:Orit}
\end{figure}


\begin{figure}[htb]
	\begin{center}
		\includegraphics[width=0.45\textwidth]{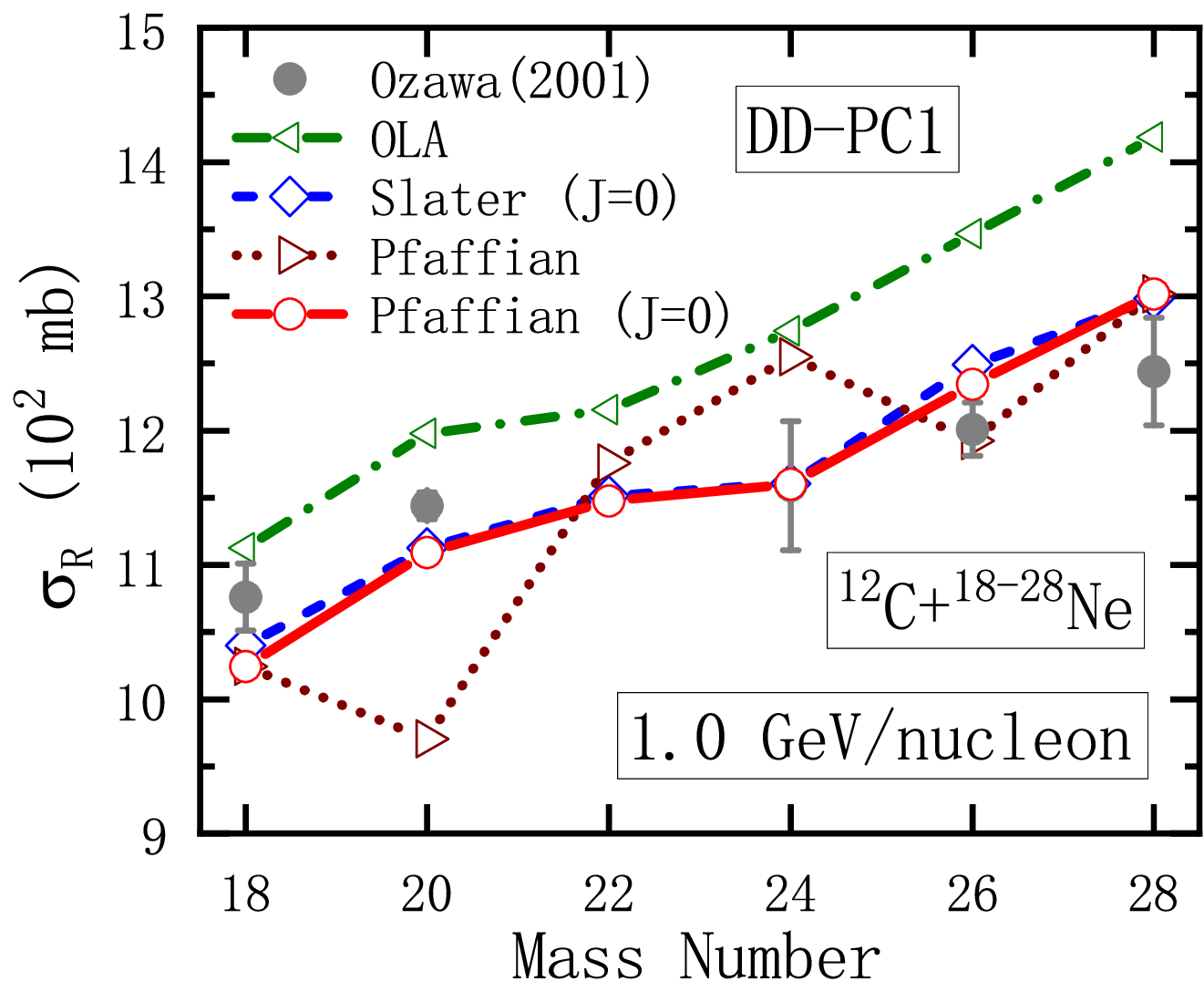}
	\end{center}
	\caption{(Color online) 
		Reaction cross sections of $^{18-28}$Ne isotopes on a $^{12}$C target at 1.0 GeV/nucleon.
		The OLA results calculated using the RHB one-body densities are shown by olive left-pointing triangles.
		The MCI results obtained using configurations sampled from the (J=0) angular-momentum-projected Pfaffian and Slater-determinant wave functions are shown by red circles and blue diamonds, respectively. For comparison, the results obtained from the fixed-orientation intrinsic Pfaffian states, with their symmetry axes aligned with the beam direction, are shown by wine-colored right-pointing triangles.
		The data (interaction cross sections) are taken from Ref.~\cite{Ozawa2001_NPA693-32} (gray dots).
	}
	\label{fig:NeCS}
\end{figure}

\begin{figure}[htb]
	\begin{center}
		\includegraphics[width=0.45\textwidth]{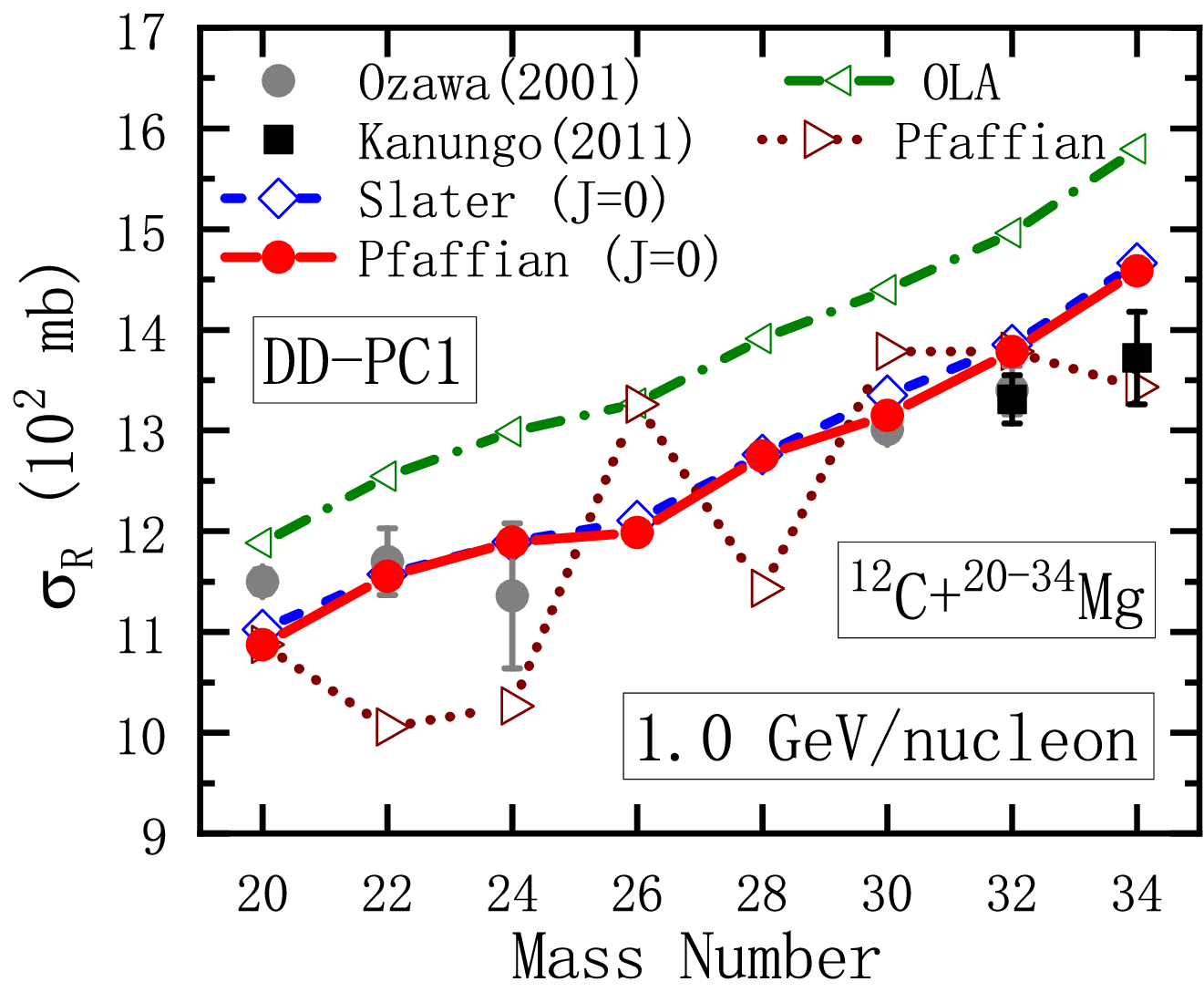}
	\end{center}
	\caption{(Color online) 
		Same as in the caption to Fig.~\ref{fig:NeCS}, but for $^{20-34}$Mg.
		The data (interaction cross sections) are taken from Ref.~\cite{Ozawa2001_NPA693-32} (gray dots) and Ref.~\cite{Kanungo2011_PRC6-061304} (black squares).
	}
	\label{fig:MgCS}
\end{figure}

As mentioned in Ref.~\cite{Nagahisa2018_PRC97-054614}, 
the root-mean-square (RMS)
radii can be used to evaluate the precision of the MCI. 
As a benchmark, the RMS radii of $^{16-24}$O isotopes 
calculated using the MCI with Pfaffian wave functions are presented in Fig.~\ref{fig:rmsR} (a).
The center-of-mass (C.M.) uncorrected RMS radii agree well with those obtained 
from RHB calculations, i.e., the deviation is at most $\sim$ 0.2\% and can be 
reduced further by increasing the number of MC configurations.
The C.M. corrected RMS radii obtained by MCI of the multidimensional integration
\begin{equation}
	\label{eq:cmR}
	\sqrt{\left\langle r^{2} \right\rangle} = 
	\sqrt{
		\frac{1}{A} \sum_{i=1}^{A} \int \cdots \int \prod_{j=1}^{A}
		d \bm{r}_{j} \left| \tilde{\bm{r}}_{j} \right|^{2} 
		\rho_{A} (\bm{R})
	 },
\end{equation}
with $\rho_{A} (\bm{R})$ as the guiding function 
are also plotted in Fig.~\ref{fig:rmsR} (a).
The C.M. corrected RMS radii are typically smaller than 
that of the corresponding C.M. uncorrected value by $\sim$ 0.05 fm, 
consistent with the trend reported 
in Ref.~\cite{Nagahisa2018_PRC97-054614}.
For the RCS of $^{12}$C+$^{16}$O, Fig.~\ref{fig:rmsR}(b)
displays the energy dependence from 40 to 1000 MeV/nucleon.
The Pfaffian-based MCI calculation shows excellent agreement with 
experimental data, particularly at higher energies where the eikonal 
approximation is most robust.
In addition, since the RHB calculations predict vanishing pairing, 
the one calculated with Pfaffian-based MCI and 
Slater determinant (consists of single-particle wave functions 
from RHB calculations but without pairing correlations) based MCI agrees. 

The interaction cross sections (ICS) $\sigma_{I}$ for the oxygen isotopic chain 
$^{16-24}$O on a $^{12}$C target at 1.0 GeV/nucleon are analyzed in Fig.~\ref{fig:Ocs}.
It is evident that the OLA, which utilizes one-body densities, 
\begin{equation}
i\chi(\bm{b}) = -\int d\bm{r} d\bm{r}^{\prime} 
\rho_{P} (\bm{r}) \rho_{T}(\bm{r}^{\prime})
\Gamma(\bm{b}+\bm{s}-\bm{s^{\prime}}),
\end{equation}
consistently overestimates the measured $\sigma_{I}$ values, 
in agreement with the results presented in Ref.~\cite{Ahmad2017_PRC95-054601}. 
In contrast, the MCI results, including both Slater determinant 
and Pfaffian wave functions-provide a significantly better 
description of the experimental data. 
This improvement demonstrates that multiple-scattering effects 
beyond the OLA are essential for a precise description of reaction 
observables in these isotopes, as already pointed out in Refs.~\cite{Suzuki2003_CRC,Ogawa1992_NPA543-722,AbuIbrahim2000_PRC62-034608,
	Takechi2009_PRC79-061601,Tran2016_PRC94-064604,
	CiofidegliAtti2011_PRC84-025205,Hatakeyama2014_JPCS569-012050}.
For $^{16-22}$O, the differences between our predictions 
and the experimental data remain within approximately 20 mb,
corresponding to at most 2\% of the measured values.
It should be noted that the predicted $\sigma_{R}$ values
are systematically smaller than the measured $\sigma_{I}$.
Strictly speaking, $\sigma_{R}$ should be larger 
than $\sigma_{I}$ because of contributions from inelastic channels, 
although the difference is expected to be negligibly small at incident energies around 1 GeV/nucleon~\cite{Ozawa2002_NPA709-60,Takechi2009_PRC79-061601,Kohama2008_PRC78-061601}. 
Therefore, the actual uncertainties of the present model may 
be slightly larger than those inferred from a direct comparison 
with the experimental $\sigma_{I}$ data.
As demonstrated in Ref.~\cite{Hatakeyama2019_NPA985-20}, 
the difference between $\sigma_{I}$ and $\sigma_{R}$
is typically no more than 2\%-3\% of $\sigma_{R}$. 
Taking this difference into account, 
our calculations underestimate the physical reaction 
cross sections by no more than approximately 5\% 
for the isotopes considered here.
Special attention should be paid to $^{24}$O where the 
present MCI underestimates the interaction cross sections 
while the OLA reproduces the experimental values fortuitously
which may indicate the need to improve of the nucleon-nucleon 
effective interactions for the study of neutron-rich nuclei in the 
mean-field approximation~\cite{Kaur2022_PRL129-142502}.

The extension to Ne and Mg requires particular care 
because their self-consistent RHB ground states are 
generally deformed. 
Fig.~\ref{fig:gsIso} shows the quadrupole deformations 
[(a)-(c)] and matter radii [(d)-(f)],
calculated with and without pairing for 
the O, Ne, and Mg chains. 
The pairing energies obtained from RHB calculations are 
also shown in Fig.~\ref{fig:gsIso} [(g)-(i)].
Whereas the oxygen ground states remain nearly spherical, 
several Ne and Mg isotopes exhibit sizable prolate or oblate deformation. 
Switching pairing on or off can also change the self-consistent equilibrium deformation. 
Consequently, a direct comparison between 
paired and unpaired intrinsic calculations does not 
isolate a pure pairing effect: 
it simultaneously changes the pairing field, 
intrinsic shape, and orientation-dependent reaction geometry.

For an axially deformed intrinsic state, 
the transverse distribution seen by the target depends 
on the angle $\beta$ between the intrinsic symmetry axis 
and the beam direction. 
This dependence is illustrated in Fig.~\ref{fig:Orit} 
for a prolate configuration of $^{20}$Ne 
and an oblate configuration of $^{24}$Ne. 
For prolate $^{20}$Ne, the cross section is smallest 
for the tip-on configuration and largest 
when the symmetry axis is perpendicular to the beam. 
The trend is reversed for oblate $^{24}$Ne. 
The intrinsic reaction cross section is therefore 
not a rotationally invariant laboratory observable.
The upper panel of Fig.~\ref{fig:Orit} displays the 
angular Gauss-Legendre weights entering the projections onto $J=0$, 2, and 4. 
The projected cross sections are calculated 
using the approximate projected wave functions 
constructed as coherent superpositions of intrinsic configurations rotated about the $y$ axis, 
as given in Eq.~\ref{eq:amp_1d}. 
In the present application, the $J=0$ projection constructs 
the appropriate laboratory-frame ground state 
and removes the arbitrary choice of intrinsic orientation.
Owing to the opposite orientation dependence of the 
reaction cross section for prolate and oblate shapes, 
the projected $J=0$, 2, and 4 states exhibit opposite spin-dependent trends. 
For the prolate nucleus $^{20}$Ne, $\sigma_R$ decreases 
with increasing $J$, 
whereas the ordering is reversed for the oblate 
nucleus $^{24}$Ne. 
Nevertheless, the differences among the projected cross 
sections remain small, at approximately 1\%.

Fig.~\ref{fig:NeCS} and~\ref{fig:MgCS} show the RCS 
for the Ne and Mg chains, respectively. 
The unprojected Pfaffian calculations correspond 
to a fixed orientation of the intrinsic symmetry axis 
relative to the beam. 
They display pronounced fluctuations that follow the 
changes in equilibrium deformation shown in Fig.~\ref{fig:gsIso}.
These fluctuations arise predominantly because differently 
deformed intrinsic states present different 
transverse geometries in the selected orientation.
After projection onto $J=0$, the calculated cross sections
exhibit much smoother isotopic behavior and generally 
follow the experimental trends. 
The projected Pfaffian and projected Slater-determinant 
results are close for most of the nuclei considered. 
This indicates that, for the present well-bound systems 
and the inclusive reaction cross section, 
the dominant differences between the original intrinsic 
calculations arise from deformation and orientation rather 
than from a strong direct sensitivity to pairing correlations.
The comparison also illustrates an important requirement 
for microscopic structure-reaction calculations. 
Although an intrinsic mean-field state provides 
a convenient representation of nuclear deformation, 
it cannot be inserted directly into a laboratory-frame reaction
calculation without specifying or restoring its orientation. 
In particular, differences between paired and unpaired 
self-consistent calculations cannot be identified with 
direct pairing effects when the two solutions have different
intrinsic deformations. 
Symmetry restoration or a physically justified treatment 
of orientation is required before such structural effects 
can be interpreted.

The calculated RCS are compared with the measured ICS. 
For the Ne isotopes, the present model underestimates 
the measured $\sigma_I$ values for $^{18}$Ne and $^{20}$Ne. 
The largest discrepancy occurs for $^{18}$Ne and amounts to approximately $4\%$.
Because the RCS is expected to exceed the ICS by approximately $2\%-3\%$,
the calculated $\sigma_R$ for $^{18}$Ne may underestimate 
the corresponding physical RCS by as much as approximately $6\%-7\%$.
By contrast, the calculations overestimate the measured $\sigma_I$
values for $^{26}$Ne and $^{28}$Ne by at most approximately $4\%$. 
Once the expected $2\%-3\%$ difference between $\sigma_I$ and $\sigma_R$
is taken into account, these results are compatible with the physical 
RCS at the $1\%-2\%$ level.

For the Mg isotopes, the calculated RCS generally 
reproduce the isotopic trend of the experimental data. 
The result for $^{20}$Mg lies below the measured $\sigma_I$ by approximately $5\%$;
after accounting for the expected difference between $\sigma_I$ and $\sigma_R$,
the physical RCS may therefore be underestimated by approximately $7\%-8\%$. 
For the heavier Mg isotopes, the calculated $\sigma_R$ values are generally 
larger than the measured $\sigma_I$ values. 
An overestimate of $\sigma_I$ by approximately $2\%-4\%$ is largely consistent 
with the expected inequality $\sigma_R>\sigma_I$. 
The largest deviation appears for $^{34}$Mg, 
for which the calculated value exceeds the measured 
interaction cross section by approximately $6\%$. 
Even after allowing for the expected $\sigma_R-\sigma_I$ difference, 
a residual discrepancy of approximately $3\%-4\%$ remains. 
Nevertheless, considering the experimental uncertainties and 
the distinction between reaction and interaction cross sections, 
the present calculations provide an overall reasonable description 
of the measured Ne and Mg systematics.

To assess the accuracy of the 1D axial projection approximation [Eq.~\ref{eq:amp_1d}], 
we have also performed a calculation using the full 2D angular integration for $^{20}$Ne 
at an incident energy of 1 GeV/nucleon. 
The resulting RCS is 1092 mb, 
compared with 1109 mb obtained using the 1D approximation at the same energy, 
corresponding to a relative difference of approximately 1.6\%. 
This small difference indicates that the 1D approximation provides a reliable 
description of the total RCS for the system considered here and does 
not affect the main conclusions of the present study.

\section{\label{sec:summary}Summary}
%
In summary, I have developed a microscopic framework that embeds 
particle-number-projected Bogoliubov many-body wave functions directly 
into the MCI of the complete Glauber multiple-scattering operator. 
The coordinate-space Pfaffian representation makes it possible to sample 
the many-body probability distribution of an RHB state without 
reducing the nuclear input to its one-body density.
Applications to O isotopes demonstrate that complete multiple-scattering  calculations 
reduce the calculated RCS relative to the OLA 
and generally improves agreement with the measured ICS. 
Applications to the deformed Ne and Mg chains show that RCS
evaluated from fixed-orientation intrinsic states can 
exhibit large geometrical variations. 
These variations are strongly correlated with the self-consistent deformation.
Restoring rotational symmetry produces smooth  $J=0$ laboratory-frame systematics 
and provides a consistent basis for comparing microscopic calculations 
with unpolarized reaction measurements.

The present work establishes the Pfaffian-based Glauber MCI 
as a general interface between Bogoliubov nuclear structure calculations 
and high-energy reaction observables. 
Future applications to weakly bound systems and more exclusive observables 
may provide direct access to correlations encoded in 
the many-body wave function beyond the one-body density.
The significant discrepancies between the calculations 
and the experimental data for nuclei at or near the drip-line, 
such as $^{24}$O, $^{18}$Ne, and $^{20}$Mg, 
may indicate limitations of the mean-field description 
in weakly bound systems, including deficiencies in the 
effective nucleon-nucleon interaction and the treatment of continuum coupling.

\begin{acknowledgments}
This work has been supported by the National Natural Science 
Foundation of China under Grant No. 12005107.
Calculations were performed on the 
Guangxi Key Laboratory of Nuclear Physics and Technology
High Performance Computing Platform.
\end{acknowledgments}


\begin{thebibliography}{46}%
	\makeatletter
	\providecommand \@ifxundefined [1]{%
		\@ifx{#1\undefined}
	}%
	\providecommand \@ifnum [1]{%
		\ifnum #1\expandafter \@firstoftwo
		\else \expandafter \@secondoftwo
		\fi
	}%
	\providecommand \@ifx [1]{%
		\ifx #1\expandafter \@firstoftwo
		\else \expandafter \@secondoftwo
		\fi
	}%
	\providecommand \natexlab [1]{#1}%
	\providecommand \enquote  [1]{``#1''}%
	\providecommand \bibnamefont  [1]{#1}%
	\providecommand \bibfnamefont [1]{#1}%
	\providecommand \citenamefont [1]{#1}%
	\providecommand \href@noop [0]{\@secondoftwo}%
	\providecommand \href [0]{\begingroup \@sanitize@url \@href}%
	\providecommand \@href[1]{\@@startlink{#1}\@@href}%
	\providecommand \@@href[1]{\endgroup#1\@@endlink}%
	\providecommand \@sanitize@url [0]{\catcode `\\12\catcode `\$12\catcode
		`\&12\catcode `\#12\catcode `\^12\catcode `\_12\catcode `\%12\relax}%
	\providecommand \@@startlink[1]{}%
	\providecommand \@@endlink[0]{}%
	\providecommand \url  [0]{\begingroup\@sanitize@url \@url }%
	\providecommand \@url [1]{\endgroup\@href {#1}{\urlprefix }}%
	\providecommand \urlprefix  [0]{URL }%
	\providecommand \Eprint [0]{\href }%
	\providecommand \doibase [0]{https://doi.org/}%
	\providecommand \selectlanguage [0]{\@gobble}%
	\providecommand \bibinfo  [0]{\@secondoftwo}%
	\providecommand \bibfield  [0]{\@secondoftwo}%
	\providecommand \translation [1]{[#1]}%
	\providecommand \BibitemOpen [0]{}%
	\providecommand \bibitemStop [0]{}%
	\providecommand \bibitemNoStop [0]{.\EOS\space}%
	\providecommand \EOS [0]{\spacefactor3000\relax}%
	\providecommand \BibitemShut  [1]{\csname bibitem#1\endcsname}%
	\let\auto@bib@innerbib\@empty
	\bibitem [{\citenamefont {Martin}\ and\ \citenamefont
		{Glauber}(1958)}]{Martin1958_PR109-1307}%
	\BibitemOpen
	\bibfield  {author} {\bibinfo {author} {\bibfnamefont {P.~C.}\ \bibnamefont
			{Martin}}\ and\ \bibinfo {author} {\bibfnamefont {R.~J.}\ \bibnamefont
			{Glauber}},\ }\bibfield  {title} {\bibinfo {title} {Relativistic theory of
			radiative orbital electron capture},\ }\href
	{https://doi.org/10.1103/physrev.109.1307} {\bibfield  {journal} {\bibinfo
			{journal} {Phys. Rev.}\ }\textbf {\bibinfo {volume} {109}},\ \bibinfo {pages}
		{1307} (\bibinfo {year} {1958})}\BibitemShut {NoStop}%
	\bibitem [{\citenamefont {Glauber}\ and\ \citenamefont
		{Matthiae}(1970)}]{Glauber1970_NPB21-135}%
	\BibitemOpen
	\bibfield  {author} {\bibinfo {author} {\bibfnamefont {R.}~\bibnamefont
			{Glauber}}\ and\ \bibinfo {author} {\bibfnamefont {G.}~\bibnamefont
			{Matthiae}},\ }\bibfield  {title} {\bibinfo {title} {High-energy scattering
			of protons by nuclei},\ }\href {https://doi.org/10.1016/0550-3213(70)90511-0}
	{\bibfield  {journal} {\bibinfo  {journal} {Nucl. Phys. B}\ }\textbf
		{\bibinfo {volume} {21}},\ \bibinfo {pages} {135} (\bibinfo {year}
		{1970})}\BibitemShut {NoStop}%
	\bibitem [{\citenamefont {d'Enterria}\ and\ \citenamefont
		{Loizides}(2021)}]{dEnterria2021_ARNPS71-315}%
	\BibitemOpen
	\bibfield  {author} {\bibinfo {author} {\bibfnamefont {D.}~\bibnamefont
			{d'Enterria}}\ and\ \bibinfo {author} {\bibfnamefont {C.}~\bibnamefont
			{Loizides}},\ }\bibfield  {title} {\bibinfo {title} {Progress in the glauber
			model at collider energies},\ }\href
	{https://doi.org/10.1146/annurev-nucl-102419-060007} {\bibfield  {journal}
		{\bibinfo  {journal} {Annu. Rev. Nucl. Part. S.}\ }\textbf {\bibinfo {volume}
			{71}},\ \bibinfo {pages} {315} (\bibinfo {year} {2021})}\BibitemShut
	{NoStop}%
	\bibitem [{\citenamefont {Horiuchi}\ \emph {et~al.}(2007)\citenamefont
		{Horiuchi}, \citenamefont {Suzuki}, \citenamefont {Abu-Ibrahim},\ and\
		\citenamefont {Kohama}}]{Horiuchi2007_PRC75-044607}%
	\BibitemOpen
	\bibfield  {author} {\bibinfo {author} {\bibfnamefont {W.}~\bibnamefont
			{Horiuchi}}, \bibinfo {author} {\bibfnamefont {Y.}~\bibnamefont {Suzuki}},
		\bibinfo {author} {\bibfnamefont {B.}~\bibnamefont {Abu-Ibrahim}},\ and\
		\bibinfo {author} {\bibfnamefont {A.}~\bibnamefont {Kohama}},\ }\bibfield
	{title} {\bibinfo {title} {Systematic analysis of reaction cross sections of
			carbon isotopes},\ }\href {https://doi.org/10.1103/physrevc.75.044607}
	{\bibfield  {journal} {\bibinfo  {journal} {Phys. Rev. C}\ }\textbf {\bibinfo
			{volume} {75}},\ \bibinfo {pages} {044607} (\bibinfo {year}
		{2007})}\BibitemShut {NoStop}%
	\bibitem [{\citenamefont {Abu-Ibrahim}\ and\ \citenamefont
		{Suzuki}(2000)}]{AbuIbrahim2000_PRC62-034608}%
	\BibitemOpen
	\bibfield  {author} {\bibinfo {author} {\bibfnamefont {B.}~\bibnamefont
			{Abu-Ibrahim}}\ and\ \bibinfo {author} {\bibfnamefont {Y.}~\bibnamefont
			{Suzuki}},\ }\bibfield  {title} {\bibinfo {title} {Scatterings of complex
			nuclei in the glauber model},\ }\href
	{https://doi.org/10.1103/physrevc.62.034608} {\bibfield  {journal} {\bibinfo
			{journal} {Phys. Rev. C}\ }\textbf {\bibinfo {volume} {62}},\ \bibinfo
		{pages} {034608} (\bibinfo {year} {2000})}\BibitemShut {NoStop}%
	\bibitem [{\citenamefont {Takechi}\ \emph {et~al.}(2009)\citenamefont
		{Takechi}, \citenamefont {Fukuda}, \citenamefont {Mihara}, \citenamefont
		{Tanaka}, \citenamefont {Chinda}, \citenamefont {Matsumasa}, \citenamefont
		{Nishimoto}, \citenamefont {Matsumiya}, \citenamefont {Nakashima},
		\citenamefont {Matsubara}, \citenamefont {Matsuta}, \citenamefont
		{Minamisono}, \citenamefont {Ohtsubo}, \citenamefont {Izumikawa},
		\citenamefont {Momota}, \citenamefont {Suzuki}, \citenamefont {Yamaguchi},
		\citenamefont {Koyama}, \citenamefont {Shinozaki}, \citenamefont {Takahashi},
		\citenamefont {Takizawa}, \citenamefont {Matsuyama}, \citenamefont
		{Nakajima}, \citenamefont {Kobayashi}, \citenamefont {Hosoi}, \citenamefont
		{Suda}, \citenamefont {Sasaki}, \citenamefont {Sato}, \citenamefont
		{Kanazawa},\ and\ \citenamefont {Kitagawa}}]{Takechi2009_PRC79-061601}%
	\BibitemOpen
	\bibfield  {author} {\bibinfo {author} {\bibfnamefont {M.}~\bibnamefont
			{Takechi}}, \bibinfo {author} {\bibfnamefont {M.}~\bibnamefont {Fukuda}},
		\bibinfo {author} {\bibfnamefont {M.}~\bibnamefont {Mihara}}, \bibinfo
		{author} {\bibfnamefont {K.}~\bibnamefont {Tanaka}}, \bibinfo {author}
		{\bibfnamefont {T.}~\bibnamefont {Chinda}}, \bibinfo {author} {\bibfnamefont
			{T.}~\bibnamefont {Matsumasa}}, \bibinfo {author} {\bibfnamefont
			{M.}~\bibnamefont {Nishimoto}}, \bibinfo {author} {\bibfnamefont
			{R.}~\bibnamefont {Matsumiya}}, \bibinfo {author} {\bibfnamefont
			{Y.}~\bibnamefont {Nakashima}}, \bibinfo {author} {\bibfnamefont
			{H.}~\bibnamefont {Matsubara}}, \bibinfo {author} {\bibfnamefont
			{K.}~\bibnamefont {Matsuta}}, \bibinfo {author} {\bibfnamefont
			{T.}~\bibnamefont {Minamisono}}, \bibinfo {author} {\bibfnamefont
			{T.}~\bibnamefont {Ohtsubo}}, \bibinfo {author} {\bibfnamefont
			{T.}~\bibnamefont {Izumikawa}}, \bibinfo {author} {\bibfnamefont
			{S.}~\bibnamefont {Momota}}, \bibinfo {author} {\bibfnamefont
			{T.}~\bibnamefont {Suzuki}}, \bibinfo {author} {\bibfnamefont
			{T.}~\bibnamefont {Yamaguchi}}, \bibinfo {author} {\bibfnamefont
			{R.}~\bibnamefont {Koyama}}, \bibinfo {author} {\bibfnamefont
			{W.}~\bibnamefont {Shinozaki}}, \bibinfo {author} {\bibfnamefont
			{M.}~\bibnamefont {Takahashi}}, \bibinfo {author} {\bibfnamefont
			{A.}~\bibnamefont {Takizawa}}, \bibinfo {author} {\bibfnamefont
			{T.}~\bibnamefont {Matsuyama}}, \bibinfo {author} {\bibfnamefont
			{S.}~\bibnamefont {Nakajima}}, \bibinfo {author} {\bibfnamefont
			{K.}~\bibnamefont {Kobayashi}}, \bibinfo {author} {\bibfnamefont
			{M.}~\bibnamefont {Hosoi}}, \bibinfo {author} {\bibfnamefont
			{T.}~\bibnamefont {Suda}}, \bibinfo {author} {\bibfnamefont {M.}~\bibnamefont
			{Sasaki}}, \bibinfo {author} {\bibfnamefont {S.}~\bibnamefont {Sato}},
		\bibinfo {author} {\bibfnamefont {M.}~\bibnamefont {Kanazawa}},\ and\
		\bibinfo {author} {\bibfnamefont {A.}~\bibnamefont {Kitagawa}},\ }\bibfield
	{title} {\bibinfo {title} {Reaction cross sections at intermediate energies
			and fermi-motion effect},\ }\href
	{https://doi.org/10.1103/physrevc.79.061601} {\bibfield  {journal} {\bibinfo
			{journal} {Phys. Rev. C}\ }\textbf {\bibinfo {volume} {79}},\ \bibinfo
		{pages} {061601} (\bibinfo {year} {2009})}\BibitemShut {NoStop}%
	\bibitem [{\citenamefont {Tran}\ \emph {et~al.}(2016)\citenamefont {Tran},
		\citenamefont {Ong}, \citenamefont {Nguyen}, \citenamefont {Tanihata},
		\citenamefont {Aoi}, \citenamefont {Ayyad}, \citenamefont {Chan},
		\citenamefont {Fukuda}, \citenamefont {Hashimoto}, \citenamefont {Hoang},
		\citenamefont {Ideguchi}, \citenamefont {Inoue}, \citenamefont {Kawabata},
		\citenamefont {Khiem}, \citenamefont {Lin}, \citenamefont {Matsuta},
		\citenamefont {Mihara}, \citenamefont {Momota}, \citenamefont {Nagae},
		\citenamefont {Nguyen}, \citenamefont {Nishimura}, \citenamefont {Ozawa},
		\citenamefont {Ren}, \citenamefont {Sakaguchi}, \citenamefont {Tanaka},
		\citenamefont {Takechi}, \citenamefont {Terashima}, \citenamefont {Wada},\
		and\ \citenamefont {Yamamoto}}]{Tran2016_PRC94-064604}%
	\BibitemOpen
	\bibfield  {author} {\bibinfo {author} {\bibfnamefont {D.~T.}\ \bibnamefont
			{Tran}}, \bibinfo {author} {\bibfnamefont {H.~J.}\ \bibnamefont {Ong}},
		\bibinfo {author} {\bibfnamefont {T.~T.}\ \bibnamefont {Nguyen}}, \bibinfo
		{author} {\bibfnamefont {I.}~\bibnamefont {Tanihata}}, \bibinfo {author}
		{\bibfnamefont {N.}~\bibnamefont {Aoi}}, \bibinfo {author} {\bibfnamefont
			{Y.}~\bibnamefont {Ayyad}}, \bibinfo {author} {\bibfnamefont {P.~Y.}\
			\bibnamefont {Chan}}, \bibinfo {author} {\bibfnamefont {M.}~\bibnamefont
			{Fukuda}}, \bibinfo {author} {\bibfnamefont {T.}~\bibnamefont {Hashimoto}},
		\bibinfo {author} {\bibfnamefont {T.~H.}\ \bibnamefont {Hoang}}, \bibinfo
		{author} {\bibfnamefont {E.}~\bibnamefont {Ideguchi}}, \bibinfo {author}
		{\bibfnamefont {A.}~\bibnamefont {Inoue}}, \bibinfo {author} {\bibfnamefont
			{T.}~\bibnamefont {Kawabata}}, \bibinfo {author} {\bibfnamefont {L.~H.}\
			\bibnamefont {Khiem}}, \bibinfo {author} {\bibfnamefont {W.~P.}\ \bibnamefont
			{Lin}}, \bibinfo {author} {\bibfnamefont {K.}~\bibnamefont {Matsuta}},
		\bibinfo {author} {\bibfnamefont {M.}~\bibnamefont {Mihara}}, \bibinfo
		{author} {\bibfnamefont {S.}~\bibnamefont {Momota}}, \bibinfo {author}
		{\bibfnamefont {D.}~\bibnamefont {Nagae}}, \bibinfo {author} {\bibfnamefont
			{N.~D.}\ \bibnamefont {Nguyen}}, \bibinfo {author} {\bibfnamefont
			{D.}~\bibnamefont {Nishimura}}, \bibinfo {author} {\bibfnamefont
			{A.}~\bibnamefont {Ozawa}}, \bibinfo {author} {\bibfnamefont {P.~P.}\
			\bibnamefont {Ren}}, \bibinfo {author} {\bibfnamefont {H.}~\bibnamefont
			{Sakaguchi}}, \bibinfo {author} {\bibfnamefont {J.}~\bibnamefont {Tanaka}},
		\bibinfo {author} {\bibfnamefont {M.}~\bibnamefont {Takechi}}, \bibinfo
		{author} {\bibfnamefont {S.}~\bibnamefont {Terashima}}, \bibinfo {author}
		{\bibfnamefont {R.}~\bibnamefont {Wada}},\ and\ \bibinfo {author}
		{\bibfnamefont {T.}~\bibnamefont {Yamamoto}},\ }\bibfield  {title} {\bibinfo
		{title} {Charge-changing cross-section measurements of $^{12-16}${C} at
			around 45{A} {MeV} and development of a glauber model for incident energies
			10{A}-2100{A} {MeV}},\ }\href {https://doi.org/10.1103/physrevc.94.064604}
	{\bibfield  {journal} {\bibinfo  {journal} {Phys. Rev. C}\ }\textbf {\bibinfo
			{volume} {94}},\ \bibinfo {pages} {064604} (\bibinfo {year}
		{2016})}\BibitemShut {NoStop}%
	\bibitem [{\citenamefont {Ciofi~degli Atti}\ \emph {et~al.}(2011)\citenamefont
		{Ciofi~degli Atti}, \citenamefont {Kopeliovich}, \citenamefont {Mezzetti},
		\citenamefont {Potashnikova},\ and\ \citenamefont
		{Schmidt}}]{CiofidegliAtti2011_PRC84-025205}%
	\BibitemOpen
	\bibfield  {author} {\bibinfo {author} {\bibfnamefont {C.}~\bibnamefont
			{Ciofi~degli Atti}}, \bibinfo {author} {\bibfnamefont {B.~Z.}\ \bibnamefont
			{Kopeliovich}}, \bibinfo {author} {\bibfnamefont {C.~B.}\ \bibnamefont
			{Mezzetti}}, \bibinfo {author} {\bibfnamefont {I.~K.}\ \bibnamefont
			{Potashnikova}},\ and\ \bibinfo {author} {\bibfnamefont {I.}~\bibnamefont
			{Schmidt}},\ }\bibfield  {title} {\bibinfo {title} {Number of collisions in
			the glauber model and beyond},\ }\href
	{https://doi.org/10.1103/physrevc.84.025205} {\bibfield  {journal} {\bibinfo
			{journal} {Phys. Rev. C}\ }\textbf {\bibinfo {volume} {84}},\ \bibinfo
		{pages} {025205} (\bibinfo {year} {2011})}\BibitemShut {NoStop}%
	\bibitem [{\citenamefont {Hatakeyama}\ \emph {et~al.}(2014)\citenamefont
		{Hatakeyama}, \citenamefont {Ebata}, \citenamefont {Horiuchi},\ and\
		\citenamefont {Kimura}}]{Hatakeyama2014_JPCS569-012050}%
	\BibitemOpen
	\bibfield  {author} {\bibinfo {author} {\bibfnamefont {S.}~\bibnamefont
			{Hatakeyama}}, \bibinfo {author} {\bibfnamefont {S.}~\bibnamefont {Ebata}},
		\bibinfo {author} {\bibfnamefont {W.}~\bibnamefont {Horiuchi}},\ and\
		\bibinfo {author} {\bibfnamefont {M.}~\bibnamefont {Kimura}},\ }\bibfield
	{title} {\bibinfo {title} {Multiple-scattering effects in proton- and
			alpha-nucleus reactions with glauber theory},\ }\href
	{https://doi.org/10.1088/1742-6596/569/1/012050} {\bibfield  {journal}
		{\bibinfo  {journal} {J. Phys.: Conf. Ser.}\ }\textbf {\bibinfo {volume}
			{569}},\ \bibinfo {pages} {012050} (\bibinfo {year} {2014})}\BibitemShut
	{NoStop}%
	\bibitem [{\citenamefont {Abu-Ibrahim}\ \emph {et~al.}(2003)\citenamefont
		{Abu-Ibrahim}, \citenamefont {Ogawa}, \citenamefont {Suzuki},\ and\
		\citenamefont {Tanihata}}]{Abu-Ibrahim2003_CPC151-369}%
	\BibitemOpen
	\bibfield  {author} {\bibinfo {author} {\bibfnamefont {B.}~\bibnamefont
			{Abu-Ibrahim}}, \bibinfo {author} {\bibfnamefont {Y.}~\bibnamefont {Ogawa}},
		\bibinfo {author} {\bibfnamefont {Y.}~\bibnamefont {Suzuki}},\ and\ \bibinfo
		{author} {\bibfnamefont {I.}~\bibnamefont {Tanihata}},\ }\bibfield  {title}
	{\bibinfo {title} {Cross section calculations in glauber model: I. core plus
			one-nucleon case},\ }\href
	{http://www.sciencedirect.com/science/article/B6TJ5-473MCCG-3/2/8097d477055e22e20d8933673d063fa5}
	{\bibfield  {journal} {\bibinfo  {journal} {Comput. Phys. Commun.}\ }\textbf
		{\bibinfo {volume} {151}},\ \bibinfo {pages} {369} (\bibinfo {year}
		{2003})}\BibitemShut {NoStop}%
	\bibitem [{\citenamefont {Sun}\ and\ \citenamefont
		{Wang}(2023)}]{Sun2023_CPC288-108726}%
	\BibitemOpen
	\bibfield  {author} {\bibinfo {author} {\bibfnamefont {Y.}~\bibnamefont
			{Sun}}\ and\ \bibinfo {author} {\bibfnamefont {S.}~\bibnamefont {Wang}},\
	}\bibfield  {title} {\bibinfo {title} {{CNOK}: A c++ glauber model code for
			single-nucleon knockout reactions},\ }\href
	{https://doi.org/10.1016/j.cpc.2023.108726} {\bibfield  {journal} {\bibinfo
			{journal} {Comput. Phys. Commun.}\ }\textbf {\bibinfo {volume} {288}},\
		\bibinfo {pages} {108726} (\bibinfo {year} {2023})}\BibitemShut {NoStop}%
	\bibitem [{\citenamefont {Kanungo}\ \emph {et~al.}(2000)\citenamefont
		{Kanungo}, \citenamefont {Tanihata}, \citenamefont {Ogawa}, \citenamefont
		{Toki},\ and\ \citenamefont {Ozawa}}]{Kanungo2000_NPA677-171}%
	\BibitemOpen
	\bibfield  {author} {\bibinfo {author} {\bibfnamefont {R.}~\bibnamefont
			{Kanungo}}, \bibinfo {author} {\bibfnamefont {I.}~\bibnamefont {Tanihata}},
		\bibinfo {author} {\bibfnamefont {Y.}~\bibnamefont {Ogawa}}, \bibinfo
		{author} {\bibfnamefont {H.}~\bibnamefont {Toki}},\ and\ \bibinfo {author}
		{\bibfnamefont {A.}~\bibnamefont {Ozawa}},\ }\bibfield  {title} {\bibinfo
		{title} {Halo structure in 19c: a glauber model analysis},\ }\href
	{https://doi.org/10.1016/s0375-9474(00)00306-7} {\bibfield  {journal}
		{\bibinfo  {journal} {Nucl. Phys. A}\ }\textbf {\bibinfo {volume} {677}},\
		\bibinfo {pages} {171} (\bibinfo {year} {2000})}\BibitemShut {NoStop}%
	\bibitem [{\citenamefont {Liu}\ \emph {et~al.}(2004)\citenamefont {Liu},
		\citenamefont {Ruan}, \citenamefont {Zhao}, \citenamefont {Zhang},
		\citenamefont {Yang}, \citenamefont {Ma}, \citenamefont {Lin}, \citenamefont
		{Chen}, \citenamefont {Wu}, \citenamefont {Zhan}, \citenamefont {Guo},
		\citenamefont {Xiao}, \citenamefont {Xu}, \citenamefont {Sun}, \citenamefont
		{Li},\ and\ \citenamefont {Chen}}]{Liu2004_PRC69-034326}%
	\BibitemOpen
	\bibfield  {author} {\bibinfo {author} {\bibfnamefont {Z.~H.}\ \bibnamefont
			{Liu}}, \bibinfo {author} {\bibfnamefont {M.}~\bibnamefont {Ruan}}, \bibinfo
		{author} {\bibfnamefont {Y.~L.}\ \bibnamefont {Zhao}}, \bibinfo {author}
		{\bibfnamefont {H.~Q.}\ \bibnamefont {Zhang}}, \bibinfo {author}
		{\bibfnamefont {F.}~\bibnamefont {Yang}}, \bibinfo {author} {\bibfnamefont
			{Z.~Y.}\ \bibnamefont {Ma}}, \bibinfo {author} {\bibfnamefont {C.~J.}\
			\bibnamefont {Lin}}, \bibinfo {author} {\bibfnamefont {B.~Q.}\ \bibnamefont
			{Chen}}, \bibinfo {author} {\bibfnamefont {Y.~W.}\ \bibnamefont {Wu}},
		\bibinfo {author} {\bibfnamefont {W.~L.}\ \bibnamefont {Zhan}}, \bibinfo
		{author} {\bibfnamefont {Z.~Y.}\ \bibnamefont {Guo}}, \bibinfo {author}
		{\bibfnamefont {G.~Q.}\ \bibnamefont {Xiao}}, \bibinfo {author}
		{\bibfnamefont {H.~S.}\ \bibnamefont {Xu}}, \bibinfo {author} {\bibfnamefont
			{Z.~Y.}\ \bibnamefont {Sun}}, \bibinfo {author} {\bibfnamefont {J.~X.}\
			\bibnamefont {Li}},\ and\ \bibinfo {author} {\bibfnamefont {Z.~J.}\
			\bibnamefont {Chen}},\ }\bibfield  {title} {\bibinfo {title} {Evidence for
			enhancement of the total reaction cross sections forp27,28with asi28target
			and examination of possibly relevant mechanisms},\ }\href
	{https://doi.org/10.1103/physrevc.69.034326} {\bibfield  {journal} {\bibinfo
			{journal} {Phys. Rev. C}\ }\textbf {\bibinfo {volume} {69}},\ \bibinfo
		{pages} {034326} (\bibinfo {year} {2004})}\BibitemShut {NoStop}%
	\bibitem [{\citenamefont {Moriguchi}\ \emph {et~al.}(2014)\citenamefont
		{Moriguchi}, \citenamefont {Ozawa}, \citenamefont {Ishimoto}, \citenamefont
		{Abe}, \citenamefont {Fukuda}, \citenamefont {Hachiuma}, \citenamefont
		{Ishibashi}, \citenamefont {Ito}, \citenamefont {Kuboki}, \citenamefont
		{Lantz}, \citenamefont {Nagae}, \citenamefont {Namihira}, \citenamefont
		{Nishimura}, \citenamefont {Ohtsubo}, \citenamefont {Ooishi}, \citenamefont
		{Suda}, \citenamefont {Suzuki}, \citenamefont {Suzuki}, \citenamefont
		{Takechi}, \citenamefont {Tanaka},\ and\ \citenamefont
		{Yamaguchi}}]{Moriguchi2014_NPA929-83}%
	\BibitemOpen
	\bibfield  {author} {\bibinfo {author} {\bibfnamefont {T.}~\bibnamefont
			{Moriguchi}}, \bibinfo {author} {\bibfnamefont {A.}~\bibnamefont {Ozawa}},
		\bibinfo {author} {\bibfnamefont {S.}~\bibnamefont {Ishimoto}}, \bibinfo
		{author} {\bibfnamefont {Y.}~\bibnamefont {Abe}}, \bibinfo {author}
		{\bibfnamefont {M.}~\bibnamefont {Fukuda}}, \bibinfo {author} {\bibfnamefont
			{I.}~\bibnamefont {Hachiuma}}, \bibinfo {author} {\bibfnamefont
			{Y.}~\bibnamefont {Ishibashi}}, \bibinfo {author} {\bibfnamefont
			{Y.}~\bibnamefont {Ito}}, \bibinfo {author} {\bibfnamefont {T.}~\bibnamefont
			{Kuboki}}, \bibinfo {author} {\bibfnamefont {M.}~\bibnamefont {Lantz}},
		\bibinfo {author} {\bibfnamefont {D.}~\bibnamefont {Nagae}}, \bibinfo
		{author} {\bibfnamefont {K.}~\bibnamefont {Namihira}}, \bibinfo {author}
		{\bibfnamefont {D.}~\bibnamefont {Nishimura}}, \bibinfo {author}
		{\bibfnamefont {T.}~\bibnamefont {Ohtsubo}}, \bibinfo {author} {\bibfnamefont
			{H.}~\bibnamefont {Ooishi}}, \bibinfo {author} {\bibfnamefont
			{T.}~\bibnamefont {Suda}}, \bibinfo {author} {\bibfnamefont {H.}~\bibnamefont
			{Suzuki}}, \bibinfo {author} {\bibfnamefont {T.}~\bibnamefont {Suzuki}},
		\bibinfo {author} {\bibfnamefont {M.}~\bibnamefont {Takechi}}, \bibinfo
		{author} {\bibfnamefont {K.}~\bibnamefont {Tanaka}},\ and\ \bibinfo {author}
		{\bibfnamefont {T.}~\bibnamefont {Yamaguchi}},\ }\bibfield  {title} {\bibinfo
		{title} {Density distribution of 14be from reaction cross-section
			measurements},\ }\href {https://doi.org/10.1016/j.nuclphysa.2014.06.003}
	{\bibfield  {journal} {\bibinfo  {journal} {Nucl. Phys. A}\ }\textbf
		{\bibinfo {volume} {929}},\ \bibinfo {pages} {83} (\bibinfo {year}
		{2014})}\BibitemShut {NoStop}%
	\bibitem [{\citenamefont {Ogawa}\ \emph {et~al.}(1992)\citenamefont {Ogawa},
		\citenamefont {Yabana},\ and\ \citenamefont {Suzuki}}]{Ogawa1992_NPA543-722}%
	\BibitemOpen
	\bibfield  {author} {\bibinfo {author} {\bibfnamefont {Y.}~\bibnamefont
			{Ogawa}}, \bibinfo {author} {\bibfnamefont {K.}~\bibnamefont {Yabana}},\ and\
		\bibinfo {author} {\bibfnamefont {Y.}~\bibnamefont {Suzuki}},\ }\bibfield
	{title} {\bibinfo {title} {Glauber model analysis of the fragmentation
			reaction cross sections of 11li},\ }\href
	{https://doi.org/10.1016/0375-9474(92)90556-y} {\bibfield  {journal}
		{\bibinfo  {journal} {Nucl. Phys. A}\ }\textbf {\bibinfo {volume} {543}},\
		\bibinfo {pages} {722} (\bibinfo {year} {1992})}\BibitemShut {NoStop}%
	\bibitem [{\citenamefont {Ozawa}\ \emph {et~al.}(2008)\citenamefont {Ozawa},
		\citenamefont {Fang}, \citenamefont {Fukuda}, \citenamefont {Iwasa},
		\citenamefont {Izumikawa}, \citenamefont {Jeppesen}, \citenamefont {Kanungo},
		\citenamefont {Koyama}, \citenamefont {Ohnishi}, \citenamefont {Ohtsubo},
		\citenamefont {Shinozaki}, \citenamefont {Suda}, \citenamefont {Suzuki},
		\citenamefont {Takahashi}, \citenamefont {Tanihata}, \citenamefont {Wu},\
		and\ \citenamefont {Yamaguchi}}]{Ozawa2008_PRC78-054313}%
	\BibitemOpen
	\bibfield  {author} {\bibinfo {author} {\bibfnamefont {A.}~\bibnamefont
			{Ozawa}}, \bibinfo {author} {\bibfnamefont {D.~Q.}\ \bibnamefont {Fang}},
		\bibinfo {author} {\bibfnamefont {M.}~\bibnamefont {Fukuda}}, \bibinfo
		{author} {\bibfnamefont {N.}~\bibnamefont {Iwasa}}, \bibinfo {author}
		{\bibfnamefont {T.}~\bibnamefont {Izumikawa}}, \bibinfo {author}
		{\bibfnamefont {H.}~\bibnamefont {Jeppesen}}, \bibinfo {author}
		{\bibfnamefont {R.}~\bibnamefont {Kanungo}}, \bibinfo {author} {\bibfnamefont
			{R.}~\bibnamefont {Koyama}}, \bibinfo {author} {\bibfnamefont
			{T.}~\bibnamefont {Ohnishi}}, \bibinfo {author} {\bibfnamefont
			{T.}~\bibnamefont {Ohtsubo}}, \bibinfo {author} {\bibfnamefont
			{W.}~\bibnamefont {Shinozaki}}, \bibinfo {author} {\bibfnamefont
			{T.}~\bibnamefont {Suda}}, \bibinfo {author} {\bibfnamefont {T.}~\bibnamefont
			{Suzuki}}, \bibinfo {author} {\bibfnamefont {M.}~\bibnamefont {Takahashi}},
		\bibinfo {author} {\bibfnamefont {I.}~\bibnamefont {Tanihata}}, \bibinfo
		{author} {\bibfnamefont {C.}~\bibnamefont {Wu}},\ and\ \bibinfo {author}
		{\bibfnamefont {Y.}~\bibnamefont {Yamaguchi}},\ }\bibfield  {title} {\bibinfo
		{title} {Measurement of the reaction cross section ofc18and observations of
			fragments fromc17andc18at80amev},\ }\href
	{https://doi.org/10.1103/physrevc.78.054313} {\bibfield  {journal} {\bibinfo
			{journal} {Phys. Rev. C}\ }\textbf {\bibinfo {volume} {78}},\ \bibinfo
		{pages} {054313} (\bibinfo {year} {2008})}\BibitemShut {NoStop}%
	\bibitem [{\citenamefont {Singh}\ \emph {et~al.}(2024)\citenamefont {Singh},
		\citenamefont {Casal}, \citenamefont {Horiuchi}, \citenamefont {Walet},\ and\
		\citenamefont {Satula}}]{Singh2024_PLB853-138694}%
	\BibitemOpen
	\bibfield  {author} {\bibinfo {author} {\bibfnamefont {J.}~\bibnamefont
			{Singh}}, \bibinfo {author} {\bibfnamefont {J.}~\bibnamefont {Casal}},
		\bibinfo {author} {\bibfnamefont {W.}~\bibnamefont {Horiuchi}}, \bibinfo
		{author} {\bibfnamefont {N.}~\bibnamefont {Walet}},\ and\ \bibinfo {author}
		{\bibfnamefont {W.}~\bibnamefont {Satula}},\ }\bibfield  {title} {\bibinfo
		{title} {Prediction of two-neutron halos in the {N}=28 isotones $^{40}${Mg}
			and $^{39}${Na}},\ }\href {https://doi.org/10.1016/j.physletb.2024.138694}
	{\bibfield  {journal} {\bibinfo  {journal} {Phys. Lett. B}\ }\textbf
		{\bibinfo {volume} {853}},\ \bibinfo {pages} {138694} (\bibinfo {year}
		{2024})}\BibitemShut {NoStop}%
	\bibitem [{\citenamefont {Varga}\ \emph {et~al.}(2002)\citenamefont {Varga},
		\citenamefont {Pieper}, \citenamefont {Suzuki},\ and\ \citenamefont
		{Wiringa}}]{Varga2002_PRC66-034611}%
	\BibitemOpen
	\bibfield  {author} {\bibinfo {author} {\bibfnamefont {K.}~\bibnamefont
			{Varga}}, \bibinfo {author} {\bibfnamefont {S.~C.}\ \bibnamefont {Pieper}},
		\bibinfo {author} {\bibfnamefont {Y.}~\bibnamefont {Suzuki}},\ and\ \bibinfo
		{author} {\bibfnamefont {R.~B.}\ \bibnamefont {Wiringa}},\ }\bibfield
	{title} {\bibinfo {title} {Monte carlo integration in glauber model analysis
			of reactions of halo nuclei},\ }\href
	{https://doi.org/10.1103/physrevc.66.034611} {\bibfield  {journal} {\bibinfo
			{journal} {Phys. Rev. C}\ }\textbf {\bibinfo {volume} {66}},\ \bibinfo
		{pages} {034611} (\bibinfo {year} {2002})}\BibitemShut {NoStop}%
	\bibitem [{\citenamefont {Nagahisa}\ and\ \citenamefont
		{Horiuchi}(2018)}]{Nagahisa2018_PRC97-054614}%
	\BibitemOpen
	\bibfield  {author} {\bibinfo {author} {\bibfnamefont {T.}~\bibnamefont
			{Nagahisa}}\ and\ \bibinfo {author} {\bibfnamefont {W.}~\bibnamefont
			{Horiuchi}},\ }\bibfield  {title} {\bibinfo {title} {Examination of the
			$^{22}${C} radius determination with interaction cross sections},\ }\href
	{https://doi.org/10.1103/physrevc.97.054614} {\bibfield  {journal} {\bibinfo
			{journal} {Phys. Rev. C}\ }\textbf {\bibinfo {volume} {97}},\ \bibinfo
		{pages} {054614} (\bibinfo {year} {2018})}\BibitemShut {NoStop}%
	\bibitem [{\citenamefont {Hatakeyama}\ and\ \citenamefont
		{Horiuchi}(2019)}]{Hatakeyama2019_NPA985-20}%
	\BibitemOpen
	\bibfield  {author} {\bibinfo {author} {\bibfnamefont {S.}~\bibnamefont
			{Hatakeyama}}\ and\ \bibinfo {author} {\bibfnamefont {W.}~\bibnamefont
			{Horiuchi}},\ }\bibfield  {title} {\bibinfo {title} {Complete glauber
			calculations for proton–nucleus inelastic cross sections},\ }\href
	{https://doi.org/10.1016/j.nuclphysa.2019.02.004} {\bibfield  {journal}
		{\bibinfo  {journal} {Nucl. Phys. A}\ }\textbf {\bibinfo {volume} {985}},\
		\bibinfo {pages} {20} (\bibinfo {year} {2019})}\BibitemShut {NoStop}%
	\bibitem [{\citenamefont {Sharma}\ \emph {et~al.}(2016)\citenamefont {Sharma},
		\citenamefont {Panda}, \citenamefont {Sharma},\ and\ \citenamefont
		{Patra}}]{Sharma2016_PRC93-014322}%
	\BibitemOpen
	\bibfield  {author} {\bibinfo {author} {\bibfnamefont {M.~K.}\ \bibnamefont
			{Sharma}}, \bibinfo {author} {\bibfnamefont {R.~N.}\ \bibnamefont {Panda}},
		\bibinfo {author} {\bibfnamefont {M.~K.}\ \bibnamefont {Sharma}},\ and\
		\bibinfo {author} {\bibfnamefont {S.~K.}\ \bibnamefont {Patra}},\ }\bibfield
	{title} {\bibinfo {title} {Search for halo structure inmg37using the glauber
			model and microscopic relativistic mean-field densities},\ }\href
	{https://doi.org/10.1103/physrevc.93.014322} {\bibfield  {journal} {\bibinfo
			{journal} {Phys. Rev. C}\ }\textbf {\bibinfo {volume} {93}},\ \bibinfo
		{pages} {014322} (\bibinfo {year} {2016})}\BibitemShut {NoStop}%
	\bibitem [{\citenamefont {Ahmad}\ \emph {et~al.}(2017)\citenamefont {Ahmad},
		\citenamefont {Usmani}, \citenamefont {Ahmad},\ and\ \citenamefont
		{Khan}}]{Ahmad2017_PRC95-054601}%
	\BibitemOpen
	\bibfield  {author} {\bibinfo {author} {\bibfnamefont {S.}~\bibnamefont
			{Ahmad}}, \bibinfo {author} {\bibfnamefont {A.~A.}\ \bibnamefont {Usmani}},
		\bibinfo {author} {\bibfnamefont {S.}~\bibnamefont {Ahmad}},\ and\ \bibinfo
		{author} {\bibfnamefont {Z.~A.}\ \bibnamefont {Khan}},\ }\bibfield  {title}
	{\bibinfo {title} {Interaction cross sections and matter radii of oxygen
			isotopes using the glauber model},\ }\href
	{https://doi.org/10.1103/physrevc.95.054601} {\bibfield  {journal} {\bibinfo
			{journal} {Phys. Rev. C}\ }\textbf {\bibinfo {volume} {95}},\ \bibinfo
		{pages} {054601} (\bibinfo {year} {2017})}\BibitemShut {NoStop}%
	\bibitem [{\citenamefont {An}\ \emph {et~al.}(2024)\citenamefont {An},
		\citenamefont {Zhang}, \citenamefont {Lu}, \citenamefont {Zhong},\ and\
		\citenamefont {Zhang}}]{An2024_PLB849-138422}%
	\BibitemOpen
	\bibfield  {author} {\bibinfo {author} {\bibfnamefont {J.-L.}\ \bibnamefont
			{An}}, \bibinfo {author} {\bibfnamefont {K.-Y.}\ \bibnamefont {Zhang}},
		\bibinfo {author} {\bibfnamefont {Q.}~\bibnamefont {Lu}}, \bibinfo {author}
		{\bibfnamefont {S.-Y.}\ \bibnamefont {Zhong}},\ and\ \bibinfo {author}
		{\bibfnamefont {S.-S.}\ \bibnamefont {Zhang}},\ }\bibfield  {title} {\bibinfo
		{title} {A unified description of the halo nucleus 37mg from microscopic
			structure to reaction observables},\ }\href
	{https://doi.org/10.1016/j.physletb.2023.138422} {\bibfield  {journal}
		{\bibinfo  {journal} {Phys. Lett. B}\ }\textbf {\bibinfo {volume} {849}},\
		\bibinfo {pages} {138422} (\bibinfo {year} {2024})}\BibitemShut {NoStop}%
	\bibitem [{\citenamefont {Rashdan}\ and\ \citenamefont
		{Sewailem}(2019)}]{Rashdan2019_PRC99-034604}%
	\BibitemOpen
	\bibfield  {author} {\bibinfo {author} {\bibfnamefont {M.}~\bibnamefont
			{Rashdan}}\ and\ \bibinfo {author} {\bibfnamefont {S.}~\bibnamefont
			{Sewailem}},\ }\bibfield  {title} {\bibinfo {title} {Effects of in-medium nn
			cross section and density distribution on the reaction cross sections of ne,
			mg, and o isotopes with c12 at 1 gev},\ }\href
	{https://doi.org/10.1103/physrevc.99.034604} {\bibfield  {journal} {\bibinfo
			{journal} {Phys. Rev. C}\ }\textbf {\bibinfo {volume} {99}},\ \bibinfo
		{pages} {034604} (\bibinfo {year} {2019})}\BibitemShut {NoStop}%
	\bibitem [{\citenamefont {Zhong}\ \emph {et~al.}(2022)\citenamefont {Zhong},
		\citenamefont {Zhang}, \citenamefont {Sun},\ and\ \citenamefont
		{Smith}}]{Zhong2022_SCPMA65-262011}%
	\BibitemOpen
	\bibfield  {author} {\bibinfo {author} {\bibfnamefont {S.-Y.}\ \bibnamefont
			{Zhong}}, \bibinfo {author} {\bibfnamefont {S.-S.}\ \bibnamefont {Zhang}},
		\bibinfo {author} {\bibfnamefont {X.-X.}\ \bibnamefont {Sun}},\ and\ \bibinfo
		{author} {\bibfnamefont {M.~S.}\ \bibnamefont {Smith}},\ }\bibfield  {title}
	{\bibinfo {title} {Study of the deformed halo nucleus $^{31}${Ne} with
			glauber model based on microscopic self-consistent structures},\ }\href
	{https://doi.org/10.1007/s11433-022-1894-6} {\bibfield  {journal} {\bibinfo
			{journal} {Science China Physics, Mechanics {\&} Astronomy}\ }\textbf
		{\bibinfo {volume} {65}},\ \bibinfo {pages} {262011} (\bibinfo {year}
		{2022})}\BibitemShut {NoStop}%
	\bibitem [{\citenamefont {Tagami}\ \emph {et~al.}(2024)\citenamefont {Tagami},
		\citenamefont {Myo},\ and\ \citenamefont
		{Yahiro}}]{Tagami2024_PRC110-064610}%
	\BibitemOpen
	\bibfield  {author} {\bibinfo {author} {\bibfnamefont {S.}~\bibnamefont
			{Tagami}}, \bibinfo {author} {\bibfnamefont {T.}~\bibnamefont {Myo}},\ and\
		\bibinfo {author} {\bibfnamefont {M.}~\bibnamefont {Yahiro}},\ }\bibfield
	{title} {\bibinfo {title} {Neutron skin thickness of pb 208 from total cross
			sections of neutron scattering at 14.137 mev and neutron skin thickness of
			ca 48 , o, n, and c isotopes from reaction and interaction cross sections},\
	}\href {https://doi.org/10.1103/physrevc.110.064610} {\bibfield  {journal}
		{\bibinfo  {journal} {Phys. Rev. C}\ }\textbf {\bibinfo {volume} {110}},\
		\bibinfo {pages} {064610} (\bibinfo {year} {2024})}\BibitemShut {NoStop}%
	\bibitem [{\citenamefont {Glauber}(1959)}]{Glauber1959_LTP1-315}%
	\BibitemOpen
	\bibfield  {author} {\bibinfo {author} {\bibfnamefont {R.~J.}\ \bibnamefont
			{Glauber}},\ }in\ \href@noop {} {\emph {\bibinfo {booktitle} {Lectures in
				Theoretical Physics}}},\ Vol.~\bibinfo {volume} {1},\ \bibinfo {editor}
	{edited by\ \bibinfo {editor} {\bibfnamefont {W.~E.}\ \bibnamefont
			{Brittion}}\ and\ \bibinfo {editor} {\bibfnamefont {L.~G.}\ \bibnamefont
			{Dunham}}}\ (\bibinfo  {publisher} {Interscience, New York},\ \bibinfo {year}
	{1959})\ p.\ \bibinfo {pages} {315}\BibitemShut {NoStop}%
	\bibitem [{\citenamefont {Abu-Ibrahim}\ \emph {et~al.}(2008)\citenamefont
		{Abu-Ibrahim}, \citenamefont {Horiuchi}, \citenamefont {Kohama},\ and\
		\citenamefont {Suzuki}}]{AbuIbrahim2008_PRC77-034607}%
	\BibitemOpen
	\bibfield  {author} {\bibinfo {author} {\bibfnamefont {B.}~\bibnamefont
			{Abu-Ibrahim}}, \bibinfo {author} {\bibfnamefont {W.}~\bibnamefont
			{Horiuchi}}, \bibinfo {author} {\bibfnamefont {A.}~\bibnamefont {Kohama}},\
		and\ \bibinfo {author} {\bibfnamefont {Y.}~\bibnamefont {Suzuki}},\
	}\bibfield  {title} {\bibinfo {title} {Reaction cross sections of carbon
			isotopes incident on a proton},\ }\href
	{https://doi.org/10.1103/physrevc.77.034607} {\bibfield  {journal} {\bibinfo
			{journal} {Phys. Rev. C}\ }\textbf {\bibinfo {volume} {77}},\ \bibinfo
		{pages} {034607} (\bibinfo {year} {2008})}\BibitemShut {NoStop}%
	\bibitem [{\citenamefont {Kaki}\ \emph {et~al.}(2012)\citenamefont {Kaki},
		\citenamefont {Suzuki},\ and\ \citenamefont
		{Wiringa}}]{Kaki2012_PRC86-044601}%
	\BibitemOpen
	\bibfield  {author} {\bibinfo {author} {\bibfnamefont {K.}~\bibnamefont
			{Kaki}}, \bibinfo {author} {\bibfnamefont {Y.}~\bibnamefont {Suzuki}},\ and\
		\bibinfo {author} {\bibfnamefont {R.~B.}\ \bibnamefont {Wiringa}},\
	}\bibfield  {title} {\bibinfo {title} {Polarized proton+4,6,8he elastic
			scattering with breakup effects in the eikonal approximation},\ }\href
	{https://doi.org/10.1103/physrevc.86.044601} {\bibfield  {journal} {\bibinfo
			{journal} {Phys. Rev. C}\ }\textbf {\bibinfo {volume} {86}},\ \bibinfo
		{pages} {044601} (\bibinfo {year} {2012})}\BibitemShut {NoStop}%
	\bibitem [{\citenamefont {Horiuchi}\ \emph
		{et~al.}(2026{\natexlab{a}})\citenamefont {Horiuchi}, \citenamefont
		{Suzuki},\ and\ \citenamefont {Wiringa}}]{Horiuchi2026_PRL136-202501}%
	\BibitemOpen
	\bibfield  {author} {\bibinfo {author} {\bibfnamefont {W.}~\bibnamefont
			{Horiuchi}}, \bibinfo {author} {\bibfnamefont {Y.}~\bibnamefont {Suzuki}},\
		and\ \bibinfo {author} {\bibfnamefont {R.}~\bibnamefont {Wiringa}},\
	}\bibfield  {title} {\bibinfo {title} {Glauber-theory calculations of
			high-energy nuclear scattering observables using variational monte carlo
			wave functions},\ }\href {https://doi.org/10.1103/ppqx-yn59} {\bibfield
		{journal} {\bibinfo  {journal} {Phys. Rev. Lett.}\ }\textbf {\bibinfo
			{volume} {136}},\ \bibinfo {pages} {202501} (\bibinfo {year}
		{2026}{\natexlab{a}})}\BibitemShut {NoStop}%
	\bibitem [{\citenamefont {Horiuchi}\ \emph
		{et~al.}(2026{\natexlab{b}})\citenamefont {Horiuchi}, \citenamefont
		{Suzuki},\ and\ \citenamefont {Wiringa}}]{Horiuchi2026_PRC113-064601}%
	\BibitemOpen
	\bibfield  {author} {\bibinfo {author} {\bibfnamefont {W.}~\bibnamefont
			{Horiuchi}}, \bibinfo {author} {\bibfnamefont {Y.}~\bibnamefont {Suzuki}},\
		and\ \bibinfo {author} {\bibfnamefont {R.~B.}\ \bibnamefont {Wiringa}},\
	}\bibfield  {title} {\bibinfo {title} {Glauber-theory analysis of nuclear
			reactions on a c 12 target with variational monte carlo wave functions},\
	}\href {https://doi.org/10.1103/gcbk-s7tc} {\bibfield  {journal} {\bibinfo
			{journal} {Phys. Rev. C}\ }\textbf {\bibinfo {volume} {113}},\ \bibinfo
		{pages} {064601} (\bibinfo {year} {2026}{\natexlab{b}})}\BibitemShut
	{NoStop}%
	\bibitem [{\citenamefont {Ring}\ and\ \citenamefont
		{Schuck}(1980)}]{Ring1980_NMBP}%
	\BibitemOpen
	\bibfield  {author} {\bibinfo {author} {\bibfnamefont {P.}~\bibnamefont
			{Ring}}\ and\ \bibinfo {author} {\bibfnamefont {P.}~\bibnamefont {Schuck}},\
	}\href@noop {} {\emph {\bibinfo {title} {The Nuclear Many Body Problem}}}\
	(\bibinfo  {publisher} {Springer, Berlin},\ \bibinfo {year}
	{1980})\BibitemShut {NoStop}%
	\bibitem [{\citenamefont {Ring}\ and\ \citenamefont
		{P.Schuck}(1980)}]{Ring1980}%
	\BibitemOpen
	\bibfield  {author} {\bibinfo {author} {\bibfnamefont {P.}~\bibnamefont
			{Ring}}\ and\ \bibinfo {author} {\bibnamefont {P.Schuck}},\ }\href@noop {}
	{\emph {\bibinfo {title} {The Nuclear Many-Body Problem}}}\ (\bibinfo
	{publisher} {Springer Verlag, New York},\ \bibinfo {year} {1980})\BibitemShut
	{NoStop}%
	\bibitem [{\citenamefont {Das}\ \emph {et~al.}(2023)\citenamefont {Das},
		\citenamefont {Majekodunmi}, \citenamefont {Biswal}, \citenamefont {Panda},\
		and\ \citenamefont {Bhuyan}}]{Das2023_NPA1037-122703}%
	\BibitemOpen
	\bibfield  {author} {\bibinfo {author} {\bibfnamefont {M.}~\bibnamefont
			{Das}}, \bibinfo {author} {\bibfnamefont {J.}~\bibnamefont {Majekodunmi}},
		\bibinfo {author} {\bibfnamefont {N.}~\bibnamefont {Biswal}}, \bibinfo
		{author} {\bibfnamefont {R.}~\bibnamefont {Panda}},\ and\ \bibinfo {author}
		{\bibfnamefont {M.}~\bibnamefont {Bhuyan}},\ }\bibfield  {title} {\bibinfo
		{title} {Correlation between the nuclear structure and reaction dynamics of
			ar-isotopes as projectile using the relativistic mean-field approach},\
	}\href {https://doi.org/10.1016/j.nuclphysa.2023.122703} {\bibfield
		{journal} {\bibinfo  {journal} {Nucl. Phys. A}\ }\textbf {\bibinfo {volume}
			{1037}},\ \bibinfo {pages} {122703} (\bibinfo {year} {2023})}\BibitemShut
	{NoStop}%
	\bibitem [{\citenamefont {Nik\v{s}i\'{c}}\ \emph {et~al.}(2011)\citenamefont
		{Nik\v{s}i\'{c}}, \citenamefont {Vretenar},\ and\ \citenamefont
		{Ring}}]{Niksic2011_PPNP66-519}%
	\BibitemOpen
	\bibfield  {author} {\bibinfo {author} {\bibfnamefont {T.}~\bibnamefont
			{Nik\v{s}i\'{c}}}, \bibinfo {author} {\bibfnamefont {D.}~\bibnamefont
			{Vretenar}},\ and\ \bibinfo {author} {\bibfnamefont {P.}~\bibnamefont
			{Ring}},\ }\bibfield  {title} {\bibinfo {title} {Relativistic nuclear energy
			density functionals: Mean-field and beyond},\ }\href
	{http://www.sciencedirect.com/science/article/pii/S0146641011000561}
	{\bibfield  {journal} {\bibinfo  {journal} {Prog. Part. Nucl. Phys.}\
		}\textbf {\bibinfo {volume} {66}},\ \bibinfo {pages} {519} (\bibinfo {year}
		{2011})}\BibitemShut {NoStop}%
	\bibitem [{\citenamefont {Nik\v{s}i\'c}\ \emph {et~al.}(2008)\citenamefont
		{Nik\v{s}i\'c}, \citenamefont {Vretenar}, \citenamefont {Lalazissis},\ and\
		\citenamefont {Ring}}]{Niksic2008_PRC77-034302}%
	\BibitemOpen
	\bibfield  {author} {\bibinfo {author} {\bibfnamefont {T.}~\bibnamefont
			{Nik\v{s}i\'c}}, \bibinfo {author} {\bibfnamefont {D.}~\bibnamefont
			{Vretenar}}, \bibinfo {author} {\bibfnamefont {G.~A.}\ \bibnamefont
			{Lalazissis}},\ and\ \bibinfo {author} {\bibfnamefont {P.}~\bibnamefont
			{Ring}},\ }\bibfield  {title} {\bibinfo {title} {Finite- to zero-range
			relativistic mean-field interactions},\ }\href
	{http://link.aps.org/doi/10.1103/PhysRevC.77.034302} {\bibfield  {journal}
		{\bibinfo  {journal} {Phys. Rev. C}\ }\textbf {\bibinfo {volume} {77}},\
		\bibinfo {pages} {034302} (\bibinfo {year} {2008})}\BibitemShut {NoStop}%
	\bibitem [{\citenamefont {Tian}\ \emph
		{et~al.}(2009{\natexlab{a}})\citenamefont {Tian}, \citenamefont {Ma},\ and\
		\citenamefont {Ring}}]{Tian2009_PLB676-44}%
	\BibitemOpen
	\bibfield  {author} {\bibinfo {author} {\bibfnamefont {Y.}~\bibnamefont
			{Tian}}, \bibinfo {author} {\bibfnamefont {Z.~Y.}\ \bibnamefont {Ma}},\ and\
		\bibinfo {author} {\bibfnamefont {P.}~\bibnamefont {Ring}},\ }\bibfield
	{title} {\bibinfo {title} {A finite range pairing force for density
			functional theory in superfluid nuclei},\ }\href
	{http://www.sciencedirect.com/science/article/pii/S0370269309004912}
	{\bibfield  {journal} {\bibinfo  {journal} {Phys. Lett. B}\ }\textbf
		{\bibinfo {volume} {676}},\ \bibinfo {pages} {44} (\bibinfo {year}
		{2009}{\natexlab{a}})}\BibitemShut {NoStop}%
	\bibitem [{\citenamefont {Tian}\ \emph
		{et~al.}(2009{\natexlab{b}})\citenamefont {Tian}, \citenamefont {Ma},\ and\
		\citenamefont {Ring}}]{Tian2009_PRC80-024313}%
	\BibitemOpen
	\bibfield  {author} {\bibinfo {author} {\bibfnamefont {Y.}~\bibnamefont
			{Tian}}, \bibinfo {author} {\bibfnamefont {Z.-Y.}\ \bibnamefont {Ma}},\ and\
		\bibinfo {author} {\bibfnamefont {P.}~\bibnamefont {Ring}},\ }\bibfield
	{title} {\bibinfo {title} {Axially deformed relativistic hartree bogoliubov
			theory with a separable pairing force},\ }\href
	{http://link.aps.org/doi/10.1103/PhysRevC.80.024313} {\bibfield  {journal}
		{\bibinfo  {journal} {Phys. Rev. C}\ }\textbf {\bibinfo {volume} {80}},\
		\bibinfo {pages} {024313} (\bibinfo {year} {2009}{\natexlab{b}})}\BibitemShut
	{NoStop}%
	\bibitem [{\citenamefont {Nik\v{s}i\'c}\ \emph {et~al.}(2010)\citenamefont
		{Nik\v{s}i\'c}, \citenamefont {Ring}, \citenamefont {Vretenar}, \citenamefont
		{Tian},\ and\ \citenamefont {Ma}}]{Niksic2010_PRC81-054318}%
	\BibitemOpen
	\bibfield  {author} {\bibinfo {author} {\bibfnamefont {T.}~\bibnamefont
			{Nik\v{s}i\'c}}, \bibinfo {author} {\bibfnamefont {P.}~\bibnamefont {Ring}},
		\bibinfo {author} {\bibfnamefont {D.}~\bibnamefont {Vretenar}}, \bibinfo
		{author} {\bibfnamefont {Y.}~\bibnamefont {Tian}},\ and\ \bibinfo {author}
		{\bibfnamefont {Z.-Y.}\ \bibnamefont {Ma}},\ }\bibfield  {title} {\bibinfo
		{title} {3d relativistic hartree-bogoliubov model with a separable pairing
			interaction: Triaxial ground-state shapes},\ }\href
	{http://link.aps.org/doi/10.1103/PhysRevC.81.054318} {\bibfield  {journal}
		{\bibinfo  {journal} {Phys. Rev. C}\ }\textbf {\bibinfo {volume} {81}},\
		\bibinfo {pages} {054318} (\bibinfo {year} {2010})}\BibitemShut {NoStop}%
	\bibitem [{\citenamefont {Webber}\ \emph {et~al.}(1990)\citenamefont {Webber},
		\citenamefont {Kish},\ and\ \citenamefont {Schrier}}]{Webber1990_PRC41-520}%
	\BibitemOpen
	\bibfield  {author} {\bibinfo {author} {\bibfnamefont {W.~R.}\ \bibnamefont
			{Webber}}, \bibinfo {author} {\bibfnamefont {J.~C.}\ \bibnamefont {Kish}},\
		and\ \bibinfo {author} {\bibfnamefont {D.~A.}\ \bibnamefont {Schrier}},\
	}\bibfield  {title} {\bibinfo {title} {Total charge and mass changing cross
			sections of relativistic nuclei in hydrogen, helium, and carbon targets},\
	}\href {https://doi.org/10.1103/physrevc.41.520} {\bibfield  {journal}
		{\bibinfo  {journal} {Phys. Rev. C}\ }\textbf {\bibinfo {volume} {41}},\
		\bibinfo {pages} {520} (\bibinfo {year} {1990})}\BibitemShut {NoStop}%
	\bibitem [{\citenamefont {Ozawa}\ \emph {et~al.}(2001)\citenamefont {Ozawa},
		\citenamefont {Suzuki},\ and\ \citenamefont
		{Tanihata}}]{Ozawa2001_NPA693-32}%
	\BibitemOpen
	\bibfield  {author} {\bibinfo {author} {\bibfnamefont {A.}~\bibnamefont
			{Ozawa}}, \bibinfo {author} {\bibfnamefont {T.}~\bibnamefont {Suzuki}},\ and\
		\bibinfo {author} {\bibfnamefont {I.}~\bibnamefont {Tanihata}},\ }\bibfield
	{title} {\bibinfo {title} {Nuclear size and related topics},\ }\href
	{https://doi.org/10.1016/s0375-9474(01)01152-6} {\bibfield  {journal}
		{\bibinfo  {journal} {Nucl. Phys. A}\ }\textbf {\bibinfo {volume} {693}},\
		\bibinfo {pages} {32} (\bibinfo {year} {2001})}\BibitemShut {NoStop}%
	\bibitem [{\citenamefont {Kanungo}\ \emph {et~al.}(2011)\citenamefont
		{Kanungo}, \citenamefont {Prochazka}, \citenamefont {Uchida}, \citenamefont
		{Horiuchi}, \citenamefont {Hagen}, \citenamefont {Papenbrock}, \citenamefont
		{Nociforo}, \citenamefont {Aumann}, \citenamefont {Boutin}, \citenamefont
		{Cortina-Gil}, \citenamefont {Davids}, \citenamefont {Diakaki}, \citenamefont
		{Farinon}, \citenamefont {Geissel}, \citenamefont {Gernhauser}, \citenamefont
		{Gerl}, \citenamefont {Janik}, \citenamefont {Jensen}, \citenamefont
		{Jonson}, \citenamefont {Kindler}, \citenamefont {Kn{\"o}bel}, \citenamefont
		{Kr{\"u}cken}, \citenamefont {Lantz}, \citenamefont {Lenske}, \citenamefont
		{Litvinov}, \citenamefont {Lommel}, \citenamefont {Mahata}, \citenamefont
		{Maierbeck}, \citenamefont {Musumarra}, \citenamefont {Nilsson},
		\citenamefont {Perro}, \citenamefont {Scheidenberger}, \citenamefont {Sitar},
		\citenamefont {Strmen}, \citenamefont {Sun}, \citenamefont {Suzuki},
		\citenamefont {Szarka}, \citenamefont {Tanihata}, \citenamefont {Weick},\
		and\ \citenamefont {Winkler}}]{Kanungo2011_PRC6-061304}%
	\BibitemOpen
	\bibfield  {author} {\bibinfo {author} {\bibfnamefont {R.}~\bibnamefont
			{Kanungo}}, \bibinfo {author} {\bibfnamefont {A.}~\bibnamefont {Prochazka}},
		\bibinfo {author} {\bibfnamefont {M.}~\bibnamefont {Uchida}}, \bibinfo
		{author} {\bibfnamefont {W.}~\bibnamefont {Horiuchi}}, \bibinfo {author}
		{\bibfnamefont {G.}~\bibnamefont {Hagen}}, \bibinfo {author} {\bibfnamefont
			{T.}~\bibnamefont {Papenbrock}}, \bibinfo {author} {\bibfnamefont
			{C.}~\bibnamefont {Nociforo}}, \bibinfo {author} {\bibfnamefont
			{T.}~\bibnamefont {Aumann}}, \bibinfo {author} {\bibfnamefont
			{D.}~\bibnamefont {Boutin}}, \bibinfo {author} {\bibfnamefont
			{D.}~\bibnamefont {Cortina-Gil}}, \bibinfo {author} {\bibfnamefont
			{B.}~\bibnamefont {Davids}}, \bibinfo {author} {\bibfnamefont
			{M.}~\bibnamefont {Diakaki}}, \bibinfo {author} {\bibfnamefont
			{F.}~\bibnamefont {Farinon}}, \bibinfo {author} {\bibfnamefont
			{H.}~\bibnamefont {Geissel}}, \bibinfo {author} {\bibfnamefont
			{R.}~\bibnamefont {Gernhauser}}, \bibinfo {author} {\bibfnamefont
			{J.}~\bibnamefont {Gerl}}, \bibinfo {author} {\bibfnamefont {R.}~\bibnamefont
			{Janik}}, \bibinfo {author} {\bibfnamefont {{\O}.}~\bibnamefont {Jensen}},
		\bibinfo {author} {\bibfnamefont {B.}~\bibnamefont {Jonson}}, \bibinfo
		{author} {\bibfnamefont {B.}~\bibnamefont {Kindler}}, \bibinfo {author}
		{\bibfnamefont {R.}~\bibnamefont {Kn{\"o}bel}}, \bibinfo {author}
		{\bibfnamefont {R.}~\bibnamefont {Kr{\"u}cken}}, \bibinfo {author}
		{\bibfnamefont {M.}~\bibnamefont {Lantz}}, \bibinfo {author} {\bibfnamefont
			{H.}~\bibnamefont {Lenske}}, \bibinfo {author} {\bibfnamefont
			{Y.}~\bibnamefont {Litvinov}}, \bibinfo {author} {\bibfnamefont
			{B.}~\bibnamefont {Lommel}}, \bibinfo {author} {\bibfnamefont
			{K.}~\bibnamefont {Mahata}}, \bibinfo {author} {\bibfnamefont
			{P.}~\bibnamefont {Maierbeck}}, \bibinfo {author} {\bibfnamefont
			{A.}~\bibnamefont {Musumarra}}, \bibinfo {author} {\bibfnamefont
			{T.}~\bibnamefont {Nilsson}}, \bibinfo {author} {\bibfnamefont
			{C.}~\bibnamefont {Perro}}, \bibinfo {author} {\bibfnamefont
			{C.}~\bibnamefont {Scheidenberger}}, \bibinfo {author} {\bibfnamefont
			{B.}~\bibnamefont {Sitar}}, \bibinfo {author} {\bibfnamefont
			{P.}~\bibnamefont {Strmen}}, \bibinfo {author} {\bibfnamefont
			{B.}~\bibnamefont {Sun}}, \bibinfo {author} {\bibfnamefont {Y.}~\bibnamefont
			{Suzuki}}, \bibinfo {author} {\bibfnamefont {I.}~\bibnamefont {Szarka}},
		\bibinfo {author} {\bibfnamefont {I.}~\bibnamefont {Tanihata}}, \bibinfo
		{author} {\bibfnamefont {H.}~\bibnamefont {Weick}},\ and\ \bibinfo {author}
		{\bibfnamefont {M.}~\bibnamefont {Winkler}},\ }\bibfield  {title} {\bibinfo
		{title} {Exploring the anomaly in the interaction cross section and matter
			radius of $^{23}${O}},\ }\href {https://doi.org/10.1103/physrevc.84.061304}
	{\bibfield  {journal} {\bibinfo  {journal} {Phys. Rev. C}\ }\textbf {\bibinfo
			{volume} {84}},\ \bibinfo {pages} {061304} (\bibinfo {year}
		{2011})}\BibitemShut {NoStop}%
	\bibitem [{\citenamefont {Yasuyuki}\ \emph {et~al.}(2003)\citenamefont
		{Yasuyuki}, \citenamefont {Kazuhiro}, \citenamefont {Rezso~G.},\ and\
		\citenamefont {Kalman}}]{Suzuki2003_CRC}%
	\BibitemOpen
	\bibfield  {author} {\bibinfo {author} {\bibfnamefont {S.}~\bibnamefont
			{Yasuyuki}}, \bibinfo {author} {\bibfnamefont {Y.}~\bibnamefont {Kazuhiro}},
		\bibinfo {author} {\bibfnamefont {L.}~\bibnamefont {Rezso~G.}},\ and\
		\bibinfo {author} {\bibfnamefont {V.}~\bibnamefont {Kalman}},\ }\href@noop {}
	{\emph {\bibinfo {title} {Structure and Reactions of Light Exotic Nuclei}}}\
	(\bibinfo  {publisher} {Taylor \& Francis, London},\ \bibinfo {year}
	{2003})\BibitemShut {NoStop}%
	\bibitem [{\citenamefont {Ozawa}\ \emph {et~al.}(2002)\citenamefont {Ozawa},
		\citenamefont {Baumann}, \citenamefont {Chulkov}, \citenamefont {Cortina},
		\citenamefont {Datta}, \citenamefont {Fernandez}, \citenamefont {Geissel},
		\citenamefont {Hammache}, \citenamefont {Itahashi}, \citenamefont {Ivanov},
		\citenamefont {Janik}, \citenamefont {Kato}, \citenamefont {Kimura},
		\citenamefont {Kobayashi}, \citenamefont {Markenroth}, \citenamefont
		{Meister}, \citenamefont {M{\"u}nzenberg}, \citenamefont {Ohtsubo},
		\citenamefont {Ohya}, \citenamefont {Okuda}, \citenamefont {Ogloblin},
		\citenamefont {Pribora}, \citenamefont {Sekiguchi}, \citenamefont
		{Sit{\'a}r}, \citenamefont {Strmen}, \citenamefont {Sugimoto}, \citenamefont
		{S{\"u}mmerer}, \citenamefont {Suzuki}, \citenamefont {Tanihata},\ and\
		\citenamefont {Yamaguchi}}]{Ozawa2002_NPA709-60}%
	\BibitemOpen
	\bibfield  {author} {\bibinfo {author} {\bibfnamefont {A.}~\bibnamefont
			{Ozawa}}, \bibinfo {author} {\bibfnamefont {T.}~\bibnamefont {Baumann}},
		\bibinfo {author} {\bibfnamefont {L.}~\bibnamefont {Chulkov}}, \bibinfo
		{author} {\bibfnamefont {D.}~\bibnamefont {Cortina}}, \bibinfo {author}
		{\bibfnamefont {U.}~\bibnamefont {Datta}}, \bibinfo {author} {\bibfnamefont
			{J.}~\bibnamefont {Fernandez}}, \bibinfo {author} {\bibfnamefont
			{H.}~\bibnamefont {Geissel}}, \bibinfo {author} {\bibfnamefont
			{F.}~\bibnamefont {Hammache}}, \bibinfo {author} {\bibfnamefont
			{K.}~\bibnamefont {Itahashi}}, \bibinfo {author} {\bibfnamefont
			{M.}~\bibnamefont {Ivanov}}, \bibinfo {author} {\bibfnamefont
			{R.}~\bibnamefont {Janik}}, \bibinfo {author} {\bibfnamefont
			{T.}~\bibnamefont {Kato}}, \bibinfo {author} {\bibfnamefont {K.}~\bibnamefont
			{Kimura}}, \bibinfo {author} {\bibfnamefont {T.}~\bibnamefont {Kobayashi}},
		\bibinfo {author} {\bibfnamefont {K.}~\bibnamefont {Markenroth}}, \bibinfo
		{author} {\bibfnamefont {M.}~\bibnamefont {Meister}}, \bibinfo {author}
		{\bibfnamefont {G.}~\bibnamefont {M{\"u}nzenberg}}, \bibinfo {author}
		{\bibfnamefont {T.}~\bibnamefont {Ohtsubo}}, \bibinfo {author} {\bibfnamefont
			{S.}~\bibnamefont {Ohya}}, \bibinfo {author} {\bibfnamefont {T.}~\bibnamefont
			{Okuda}}, \bibinfo {author} {\bibfnamefont {A.}~\bibnamefont {Ogloblin}},
		\bibinfo {author} {\bibfnamefont {V.}~\bibnamefont {Pribora}}, \bibinfo
		{author} {\bibfnamefont {M.}~\bibnamefont {Sekiguchi}}, \bibinfo {author}
		{\bibfnamefont {B.}~\bibnamefont {Sit{\'a}r}}, \bibinfo {author}
		{\bibfnamefont {P.}~\bibnamefont {Strmen}}, \bibinfo {author} {\bibfnamefont
			{S.}~\bibnamefont {Sugimoto}}, \bibinfo {author} {\bibfnamefont
			{K.}~\bibnamefont {S{\"u}mmerer}}, \bibinfo {author} {\bibfnamefont
			{T.}~\bibnamefont {Suzuki}}, \bibinfo {author} {\bibfnamefont
			{I.}~\bibnamefont {Tanihata}},\ and\ \bibinfo {author} {\bibfnamefont
			{Y.}~\bibnamefont {Yamaguchi}},\ }\bibfield  {title} {\bibinfo {title}
		{Measurements of the interaction cross sections for ar and cl isotopes},\
	}\href {https://doi.org/10.1016/s0375-9474(02)01071-0} {\bibfield  {journal}
		{\bibinfo  {journal} {Nucl. Phys. A}\ }\textbf {\bibinfo {volume} {709}},\
		\bibinfo {pages} {60} (\bibinfo {year} {2002})}\BibitemShut {NoStop}%
	\bibitem [{\citenamefont {Kohama}\ \emph {et~al.}(2008)\citenamefont {Kohama},
		\citenamefont {Iida},\ and\ \citenamefont
		{Oyamatsu}}]{Kohama2008_PRC78-061601}%
	\BibitemOpen
	\bibfield  {author} {\bibinfo {author} {\bibfnamefont {A.}~\bibnamefont
			{Kohama}}, \bibinfo {author} {\bibfnamefont {K.}~\bibnamefont {Iida}},\ and\
		\bibinfo {author} {\bibfnamefont {K.}~\bibnamefont {Oyamatsu}},\ }\bibfield
	{title} {\bibinfo {title} {Difference between interaction cross sections and
			reaction cross sections},\ }\href
	{https://doi.org/10.1103/physrevc.78.061601} {\bibfield  {journal} {\bibinfo
			{journal} {Phys. Rev. C}\ }\textbf {\bibinfo {volume} {78}},\ \bibinfo
		{pages} {061601} (\bibinfo {year} {2008})}\BibitemShut {NoStop}%
	\bibitem [{\citenamefont {Kaur}\ \emph {et~al.}(2022)\citenamefont {Kaur},
		\citenamefont {Kanungo}, \citenamefont {Horiuchi}, \citenamefont {Hagen},
		\citenamefont {Holt}, \citenamefont {Hu}, \citenamefont {Miyagi},
		\citenamefont {Suzuki}, \citenamefont {Ameil}, \citenamefont {Atkinson},
		\citenamefont {Ayyad}, \citenamefont {Bagchi}, \citenamefont {Cortina-Gil},
		\citenamefont {Dillmann}, \citenamefont {Estrad{\'e}}, \citenamefont
		{Evdokimov}, \citenamefont {Farinon}, \citenamefont {Geissel}, \citenamefont
		{Guastalla}, \citenamefont {Janik}, \citenamefont {Kn{\"o}bel}, \citenamefont
		{Kurcewicz}, \citenamefont {Litvinov}, \citenamefont {Marta}, \citenamefont
		{Mostazo}, \citenamefont {Mukha}, \citenamefont {Nociforo}, \citenamefont
		{Ong}, \citenamefont {Otsuka}, \citenamefont {Pietri}, \citenamefont
		{Prochazka}, \citenamefont {Scheidenberger}, \citenamefont {Sitar},
		\citenamefont {Strmen}, \citenamefont {Takechi}, \citenamefont {Tanaka},
		\citenamefont {Tanihata}, \citenamefont {Terashima}, \citenamefont {Vargas},
		\citenamefont {Weick},\ and\ \citenamefont
		{Winfield}}]{Kaur2022_PRL129-142502}%
	\BibitemOpen
	\bibfield  {author} {\bibinfo {author} {\bibfnamefont {S.}~\bibnamefont
			{Kaur}}, \bibinfo {author} {\bibfnamefont {R.}~\bibnamefont {Kanungo}},
		\bibinfo {author} {\bibfnamefont {W.}~\bibnamefont {Horiuchi}}, \bibinfo
		{author} {\bibfnamefont {G.}~\bibnamefont {Hagen}}, \bibinfo {author}
		{\bibfnamefont {J.}~\bibnamefont {Holt}}, \bibinfo {author} {\bibfnamefont
			{B.}~\bibnamefont {Hu}}, \bibinfo {author} {\bibfnamefont {T.}~\bibnamefont
			{Miyagi}}, \bibinfo {author} {\bibfnamefont {T.}~\bibnamefont {Suzuki}},
		\bibinfo {author} {\bibfnamefont {F.}~\bibnamefont {Ameil}}, \bibinfo
		{author} {\bibfnamefont {J.}~\bibnamefont {Atkinson}}, \bibinfo {author}
		{\bibfnamefont {Y.}~\bibnamefont {Ayyad}}, \bibinfo {author} {\bibfnamefont
			{S.}~\bibnamefont {Bagchi}}, \bibinfo {author} {\bibfnamefont
			{D.}~\bibnamefont {Cortina-Gil}}, \bibinfo {author} {\bibfnamefont
			{I.}~\bibnamefont {Dillmann}}, \bibinfo {author} {\bibfnamefont
			{A.}~\bibnamefont {Estrad{\'e}}}, \bibinfo {author} {\bibfnamefont
			{A.}~\bibnamefont {Evdokimov}}, \bibinfo {author} {\bibfnamefont
			{F.}~\bibnamefont {Farinon}}, \bibinfo {author} {\bibfnamefont
			{H.}~\bibnamefont {Geissel}}, \bibinfo {author} {\bibfnamefont
			{G.}~\bibnamefont {Guastalla}}, \bibinfo {author} {\bibfnamefont
			{R.}~\bibnamefont {Janik}}, \bibinfo {author} {\bibfnamefont
			{R.}~\bibnamefont {Kn{\"o}bel}}, \bibinfo {author} {\bibfnamefont
			{J.}~\bibnamefont {Kurcewicz}}, \bibinfo {author} {\bibfnamefont
			{Y.}~\bibnamefont {Litvinov}}, \bibinfo {author} {\bibfnamefont
			{M.}~\bibnamefont {Marta}}, \bibinfo {author} {\bibfnamefont
			{M.}~\bibnamefont {Mostazo}}, \bibinfo {author} {\bibfnamefont
			{I.}~\bibnamefont {Mukha}}, \bibinfo {author} {\bibfnamefont
			{C.}~\bibnamefont {Nociforo}}, \bibinfo {author} {\bibfnamefont
			{H.}~\bibnamefont {Ong}}, \bibinfo {author} {\bibfnamefont {T.}~\bibnamefont
			{Otsuka}}, \bibinfo {author} {\bibfnamefont {S.}~\bibnamefont {Pietri}},
		\bibinfo {author} {\bibfnamefont {A.}~\bibnamefont {Prochazka}}, \bibinfo
		{author} {\bibfnamefont {C.}~\bibnamefont {Scheidenberger}}, \bibinfo
		{author} {\bibfnamefont {B.}~\bibnamefont {Sitar}}, \bibinfo {author}
		{\bibfnamefont {P.}~\bibnamefont {Strmen}}, \bibinfo {author} {\bibfnamefont
			{M.}~\bibnamefont {Takechi}}, \bibinfo {author} {\bibfnamefont
			{J.}~\bibnamefont {Tanaka}}, \bibinfo {author} {\bibfnamefont
			{I.}~\bibnamefont {Tanihata}}, \bibinfo {author} {\bibfnamefont
			{S.}~\bibnamefont {Terashima}}, \bibinfo {author} {\bibfnamefont
			{J.}~\bibnamefont {Vargas}}, \bibinfo {author} {\bibfnamefont
			{H.}~\bibnamefont {Weick}},\ and\ \bibinfo {author} {\bibfnamefont
			{J.}~\bibnamefont {Winfield}},\ }\bibfield  {title} {\bibinfo {title} {Proton
			distribution radii of $^{16-24}$o : Signatures of new shell closures and
			neutron skin},\ }\href {https://doi.org/10.1103/physrevlett.129.142502}
	{\bibfield  {journal} {\bibinfo  {journal} {Phys. Rev. Lett.}\ }\textbf
		{\bibinfo {volume} {129}},\ \bibinfo {pages} {142502} (\bibinfo {year}
		{2022})}\BibitemShut {NoStop}%
\end{thebibliography}

%

\end{document}